\documentclass[useAMS,usenatbib]{mn2e}
\usepackage{graphicx}

\newcommand{\OO}{_{_{\!0\!}}}

\title[The fragmentation of expanding shells II: Thickness matters]{
The fragmentation of expanding shells II:\\
Thickness matters.}

\author[Richard~W\"unsch, James E. Dale, Jan Palou\v{s}, Anthony P. Whitworth]
{R.~W\"unsch$^{1,2}$,\thanks{E-mail: richard@wunsch.cz}
J.~E.~Dale$^{1}$, J.~Palou\v{s}$^{1}$ and A.~P.~Whitworth$^{2}$\\
$^{1}$Astronomical Institute, Academy of Sciences of the Czech
Republic, Bo\v{c}n\'\i\ II 1401, 141 31 Prague, Czech Republic\\
$^{2}$School of Physics and Astronomy, Cardiff University, Queens Buildings, The Parade, Cardiff, CF24 3AA}
\begin{document}

\date{Received: }

\pagerange{\pageref{firstpage}--\pageref{lastpage}} \pubyear{2009}

\maketitle

\label{firstpage}

\def\mnras{MNRAS}
\def\apj{ApJ}
\def\aap{A\&A}
\def\apjl{ApJL}
\def\apjs{ApJS}
\def\bain{BAIN}
\def\araa{ARA\&A}
\def\pasp{PASP}
\def\pasj{PASJ}
\def\aj{AJ}
\def\ga{\sim}

\begin{abstract} 
We study analytically the development of gravitational instability in
an expanding shell having finite thickness. We consider three models for the
radial density profile of the shell: (i) an analytic uniform-density model, (ii)
a semi-analytic model obtained by numerical solution of the hydrostatic
equilibrium equation, and (iii) a 3D hydrodynamic simulation. We show that all
three profiles are in close agreement, and this allows us to use the first model
to describe fragments in the radial direction of the shell. We then use
non-linear equations describing the time-evolution of a uniform oblate spheroid
to derive the growth rates of shell fragments having different sizes. This
yields a dispersion relation which depends on the shell thickness, and hence on
the pressure confining the shell. We compare this dispersion relation with the
dispersion relation obtained using the standard thin-shell analysis, and show
that, if the confining pressure is low, only large fragments are unstable. On
the other hand, if the confining pressure is high, fragments smaller than
predicted by the thin-shell analysis become unstable. Finally, we compare the
new dispersion relation with the results of 3D hydrodynamic simulations,
and show that the two are in good agreement. 
\end{abstract}

\begin{keywords}
stars: formation, ISM: HII regions
\end{keywords}

\section{Introduction} 

Shells and bubbles are common morphological features in the interstellar media
of galaxies, and have been observed at many different wavelengths. For instance,
\citet{2002ApJ...578..176M} and \citet{2005A&A...437..101E} have identified
$\sim 1000$ expanding shells in HI surveys of the Milky Way. HI shells have also
been found in the LMC \citep{1998ApJ...503..674K}, the SMC
\citep{2005MNRAS.360.1171H}, and other nearby galaxies. In the infrared,
\citet{2006ApJ...649..759C,2007ApJ...670..428C} have assembled a catalogue of
$\sim 600$ shells found in the GLIMPSE survey of our Galaxy, using the SPITZER
space telescope. Shells have been detected in the Wisconsin H$\alpha$ mapper
data \citep[WHAM;][]{2003ApJS..149..405H}. Shells have also been studied in
millimetre molecular line emission, radio continuum and X-rays.

It is generally believed that a substantial fraction of these shells is formed
by feedback from massive stars. In addition, it has been suggested by
\citet{1977ApJ...214..725E} that expanding shells can trigger star formation by
the {\it collect-and-collapse}  mechanism. In this mechanism, a massive star (or
a group of stars) injects energy into the interstellar medium in the form of
stellar winds and ionising radiation, and creates a bubble of gas with high
temperature. The bubble expands due to its high internal pressure, and sweeps up
the ambient medium into a dense shell. The shell cools down, becomes
gravitationally unstable, fragments and forms a new generation of stars. If the
new generation includes massive stars, the whole process may repeat itself,
leading to sequential self-propagating star formation. Various modifications to
the collect-and-collapse mechanism are reviewed by \citet{1998ASPC..148..150E}.

The collect-and-collapse mechanism has been tested observationally in a series
of papers by \citet{2003A&A...408L..25D, 2005A&A...433..565D,
2006A&A...458..191D, 2008A&A...482..585D, 2009A&A...496..177D} and
\citet{2006A&A...446..171Z}. Using data at various wavelengths, they identify
and study several objects in which star formation appears to have been triggered
at the borders of HII regions. \citet{2008ApJ...681.1341W,2009ApJ...694..546W}
analyse six shells from the GLIMPSE survey, and determine the properties of
young stellar objects concentrated in these shells.

The gravitational instability of an expanding infinitesimally thin shell has
been studied theoretically by several authors. \citet{1983ApJ...274..152V}
derives a dispersion relation (perturbation growth rate as a function of
wavenumber) for an infinitesimally thin shell, using decomposition into
spherical harmonics. A very similar dispersion relation is obtained by
\citet{1994ApJ...427..384E} using perturbation analysis of linearised
two-dimensional hydrodynamic equations. \citet{1994MNRAS.268..291W} use the
equation of motion for a two-dimensional fragment forming on the surface of a
shell, and find expressions for the size of the first and most unstable
fragment, and for the time at which the instability starts. All these analyses
use a very similar physical model (an infinitesimally thin shell which expands
into a uniform medium), and although the mathematical description in each of
them is very different, the results regarding fragment sizes and growth rates
are very similar.

We have tested these analytic predictions for the growth rate of gravitational
instability, based on the thin-shell approximation, in \citet[][hereafter 
Paper~I]{2009MNRAS.398.1537D}. We use two very different three-dimensional
hydrodynamic codes (an Eulerian AMR code and a Lagrangian SPH code) to simulate
the evolution and subsequent fragmentation of expanding self-gravitating shells.
Our setup differs slightly from the one studied by Vishniac, Elmegreen and
Whitworth, in that our shell does not accrete mass from the ambient medium. This
modification requires only minor changes to the linear theory of the thin shell
gravitational instability, which we describe in \S\ref{sec:thin_shell}.

We make this modification because an accreting shell is prone to the Vishniac
dynamical instability \citep{1983ApJ...274..152V}, which would grow quickly and
make an analysis of gravitational instability impossible. On the other hand, we
want to have the option to keep the shell thin, in order to test the thin-shell
approximation. Therefore, we fill the shell interior and exterior with a hot
rarefied gas and the pressure of this gas confines the shell. However, because
of its low density, the ram pressure of the accreted gas, which is crucial for
the development of the Vishniac instability, is negligible. We also keep the
pressure of the rarefied gas constant throughout the evolution, so that the
shell is effectively in free fall, and development of the Rayleigh-Taylor
instability is suppressed. 

In Paper~I, we find  excellent agreement between the two numerical codes.
However, for simulations with low confining pressure, in which the shell becomes
thick, the simulations do not agree well with the predictions of the thin-shell
approximation. The agreement is better for simulations in which the confining
pressure is chosen so that the shell thickness stays approximately constant
during its evolution.

In this paper (Paper~II) we study the gravitational instability of a thick
shell analytically, in order to understand how a thick shell fragments, and why
the fragmentation differs from the predictions of the thin-shell approximation.
We show that the confining pressure, which defines the shell thickness, is an
important factor in determining the range of unstable wavelengths. Therefore, we
call the new model {\em pressure assisted gravitational instability} (PAGI).

The outline of the paper is as follows. In \S\ref{sec:radprof} we derive a
semi-analytic description of the equilibrium radial profile of the shell, and
compare it with our 3D numerical simulations, and with the simple
uniform-density model of the shell that we adopt in the subsequent analysis.
Section \S\ref{sec:gi} deals in detail with modelling gravitational instability
in a shell. We briefly derive the dispersion relation for a non-acreting shell,
using the thin-shell approximation; we give a description of a fragment forming
in a shell with non-zero thickness, and compare it with the description used in
the thin-shell approximation; finally we derive a new dispersion relation for a
thick pressure-confined shell. In \S\ref{sec:num}, we compare the new dispersion
relation with growth rates measured in our 3D numerical simulations, in
\S\ref{discussion} we discuss the astrophysical consequences of our results, and
in \S\ref{conclusions} we summarise our conclusions.


\section{Radial density profile of the isothermal shell}\label{sec:radprof} 

Since we wish to study the influence of the shell's thickness on its
gravitational instability, it is essential first to understand the shell's
equilibrium radial structure. In Paper~I we set the initial radial density
profile of the shell using a simple local model for the hydrostatic equilibrium
of the gas layer, and assuming that the shell radius is large compared with its
thickness: 
\begin{equation}\label{sech2}
\rho(R)=\rho_{_{\rm O}}\,{\rm sech}^{2}\left[\left(\frac{2\pi G\rho_{_{\rm O}} }{c_{_{\rm
S}}^{2}}\right)^{\frac{1}{2}}(R-R_{_{\rm O}})\right]\,. 
\end{equation}
Here $R$ is the radial coordinate, 
$\rho_{_{\rm O}}$ is the maximum density, which appears at the radius
$R_{_{\rm O}}$, and $c_{_{\rm S}}$ is the isothermal sound-speed.

A more realistic density profile can be obtained by dropping the assumption that
the density profile is symmetric about $R_{_{\rm O}}$ and solving the
equation of hydrostatic equilibrium in the form 
\begin{eqnarray}\label{he}
c_{_{\rm S}}^2\frac{d\rho}{dR}&=&\left\{-\,\frac{GM(R)}{R^2}\,+
\,\frac{\alpha GM_{_{\rm TOT}} }{R_{\!_\alpha}^2}\right\}\rho(R)\,.
\end{eqnarray}
Here $M(R)$ is the mass within radius $R$, and $\alpha$ is a dimensionless
number between $0$ and $1$ which has to be found numerically (see below). The
first term on the RHS of Eqn. \ref{he} represents the net gravitational
acceleration at radius $R_{\!_\alpha}$. The second term on the RHS represents the
gravitational acceleration of the material at radius $R_{\!_\alpha}$, which is
the radius that contains a fraction $\alpha$ of the total mass of the shell.
This gravitational acceleration is subtracted from the net gravitational
acceleration, on the grounds that it contributes to the bulk deceleration of the
centre of mass of the expanding shell, rather than its radial profile.

Switching to a Lagrangian formalism (i.e. using $M$ as the independent variable)
Eqn. \ref{he} becomes
\begin{eqnarray}
c_{_{\rm S}}^2\left\{\frac{d^2R/dM^2}{(dR/dM)^2}\,+\,\frac{2}{R(M)}\right\}& 
= &\frac{GM}{R^2(M)}\,-\,\frac{\alpha GM_{_{\rm TOT} }}{R_{\!_\alpha}^2}\,.
\end{eqnarray}
Introducing dimensionless variables,
\begin{eqnarray}
\xi&=&\frac{R}{R_{\!_\alpha}}\,,\\
\mu&=&\frac{M(R)}{M_{_{\rm TOT}} }\,,
\end{eqnarray}
this reduces to
\begin{eqnarray}
\label{eq:ximu}
\frac{\xi''}{(\xi')^2}\,+\,\frac{2}{\xi}&=&\frac{GM_{_{\rm TOT}} }{R_{\!_\alpha}
c_{_{\rm S}}^2}\left\{\frac{\mu}{\xi^2}\,-\,\alpha\right\}\,,
\end{eqnarray}
where $\xi'\equiv d\xi/d\mu$ and $\xi''\equiv d^2\xi/d\mu^2$.

The shell density profile can be found by solving Eqn. (\ref{eq:ximu})
numerically, but we also need to determine the appropriate boundary conditions
and the value of $\alpha$. The shell is confined by the external pressure
$P_{_{\rm EXT}}$ on both its inner and outer surfaces, and therefore the density
at these surfaces has to be $\rho_{_{\rm B}} = P_{_{\rm EXT}}/c_{_{\rm S}}^2$.
Inserting this into the mass conservation law we obtain the inner boundary
condition in the form $\left(\xi'\xi^2\right)_{_{\mu=0}} = M_{_{\rm TOT}}c_{_{\rm S}}^2/(4\pi
R_{\!_\alpha}^3P_{_{\rm EXT}})$. The value $\xi(\mu=0)$ is unknown, but it can be found
by iterating until the boundary condition is also fulfilled at the
outer surface, i.e. $\left(\xi'\xi^2\right)_{_{\mu=1}} = M_{_{\rm TOT}}c_{_{\rm S}}^2/(4\pi
R_{\!_\alpha}^3P_{_{\rm EXT}})$. The parameter $\alpha$ can be found by adding another
level of iteration in which $\alpha$ is varied until it fulfils the initial
assumption $\xi(\mu = \alpha) = 1$.

\begin{figure}
\includegraphics[width=8cm]{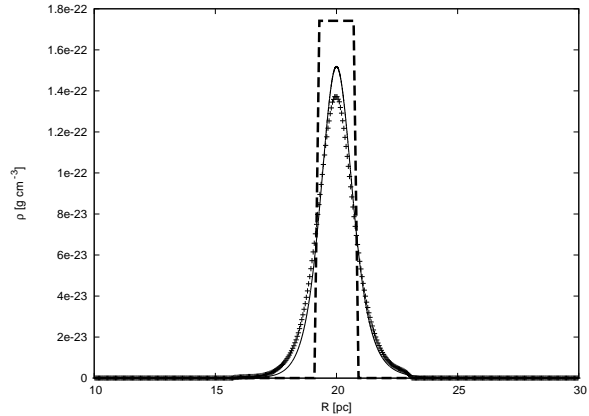}
\caption{Radial profile of the shell at $t = 7.2$~Myr in 3D AMR simulation (plus
symbols), semi-analytical model (solid line) and uniform shell with the
same column density (dashed line).} 
\label{fig:radprof}
\end{figure}

\begin{figure}
\includegraphics[width=8cm]{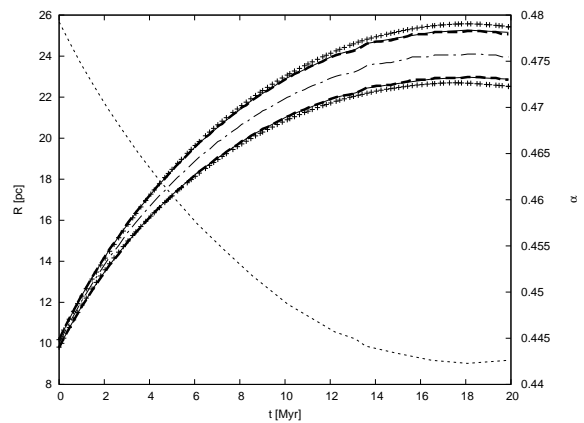}
\caption{Evolution of the shell FWHM. The dash-dotted line denotes the
position of the shell peak density, the other lines below and above it show
positions of the the shell inner and outer edge measured at half of the peak
density for the three models: 3D AMR simulation (plus symbols), semi-analytical
model (thin solid line) and uniform shell (thick dashed line). The thin dashed
line shows the evolution of the parameter $\alpha$.}
\label{fig:radevol}
\end{figure}

Figure \ref{fig:radprof} compares the radial density profile of the shell in the
AMR run from Paper~I with $P_{_{\rm EXT}} = 10^{-17}$~dyne\,cm$^{-2}$, at
time $7.2\,{\rm Myr}$, with the semi-analytic profile obtained using Eqn.
(\ref{eq:ximu}) and the same shell parameters. Both profiles agree very well
with each other, which means that the assumption of hydrostatic equilibrium is
correct and that the shell is well resolved in the numerical simulation. The
time evolution of the shell radius and its FWHM are shown in
Figure~\ref{fig:radevol}, and this also demonstrates excellent agreement between
the numerical and semi-analytic treatments. In \S\ref{sec:gi} we assume that the
shell has uniform density with half-thickness given by
\begin{equation}\label{zus}
z_{_{\rm US}} = \frac{\Sigma_{_{\rm O}} c_{_{\rm S}}^2}{2P_{_{\rm EXT}}+\pi G \Sigma_{_{\rm O}}^2}
\end{equation}
where $\Sigma_{_{\rm O}}\equiv M_{_{\rm TOT}}/(4\pi R_{\!_\alpha}^2)$ is the shell
surface density. This equation is obtained by splitting the shell into two
identical layers and then equating the external pressure force and gravitational
force acting on one layer due to the other one with the internal pressure force
in between the layers. Figures \ref{fig:radprof} and \ref{fig:radevol} also
display the density profile, mean radius and FWHM of this uniform-density shell
model, and show that it is a good approximation to the more detailed models.
The agreement between the numerical, semi-analytic and uniform-density models is
even better for runs with higher external pressures $P_{_{\rm EXT}} =
10^{-13}$ and $P_{_{\rm EXT}} = 5\times 10^{-13}$~dyne\,cm$^{-2}$.

Figure~\ref{fig:radevol} also shows the evolution of the parameter $\alpha$. It
is close to $0.5$ in the beginning of the expansion when the shell is thin and
its profile is symmetric, and only decreases by $\sim 10\%$ subsequently.
This result supports the assumption of $\alpha = 0.5$ made by
\citet{2002MNRAS.329..641W}.


\section{Growth rate of the gravitational instability}\label{sec:gi} 

The main aim of this paper is to derive the dispersion relation for
gravitational instability of a thick shell, i.e. perturbation growth rate as a
function of perturbation wavenumber. For this purpose we model a forming
fragment as a uniform-density spheroid whose radial excursions are
described by non-linear differential equations. \S\ref{sec:thin_shell} briefly summarises the derivation of the thin-shell dispersion relation for the case of a non-accreting shell. \S\ref{sec:hore} describes the spheroidal model for a fragment forming in a thick shell, and explains in detail the differences between this model and the model underlying the thin-shell approximation. \S\ref{sec:disprel} presents the derivation of the PAGI dispersion relation and compares it with the thin-shell dispersion relation.

\subsection{Thin shell approximation}\label{sec:thin_shell}

Consider an infinitesimally thin shell with radius $R$ and surface density
$\Sigma$, which expands with velocity $V$ into a vacuum. The shell is
isothermal, with sound speed $c_{_{\rm S}}$. The equations describing transverse
flows of gas inside the shell are\footnote{In \citet{1994ApJ...427..384E} and
\citet{2001A&A...374..746W} the shell was assumed to accrete mass from the
ambient uniform medium, and this resulted in a factor of $4$ in front of the
last term in the equation of motion, Eqn. \ref{hde:eom}}.
\begin{equation}\label{hde:ce}
\frac{\partial \Sigma}{\partial t}+R\Sigma\nabla_\mathrm{T}\cdot\Omega+2\Sigma\frac{V}{R}=0\,,
\end{equation}
\begin{equation}\label{hde:eom}
\Sigma R\frac{\partial\Omega}{\partial t}=-c_{_{\rm
S}}^2\nabla_\mathrm{T}\Sigma-\Sigma\nabla\Phi-\Sigma V \Omega\,,
\end{equation}
\begin{equation}\label{hde:pe}
\nabla^2\Phi=4\pi G\Sigma\delta(r-R)\,.
\end{equation}
Here $\Omega$ is the angular velocity of transverse flows, $\Phi$ is the gravitational potential, and $\nabla_{_{\!\rm T}}$ and $\nabla_{_{\!\rm T}}\cdot$ denote the two-dimensional gradient and divergence, respectively. The instability growth rate is obtained by substituting plane-wave perturbations parametrised by the dimensionless wavenumber $l = kR$ (see \citealt{1994ApJ...427..384E}, \citealt{2001A&A...374..746W} and Paper~I for details),
\begin{eqnarray}\label{pert}
\Sigma & = & \Sigma_{_{\rm O}} + \Sigma_l\cos(l\theta)\,, \nonumber\\
\Omega & = & \Omega_l\sin(l\theta)\,, \\
\Phi & = & \Phi_{_{\rm O}} + \Phi_l\cos(l\theta)\,. \nonumber
\end{eqnarray}
Here $\Sigma_{_{\rm O}}$ and $\Phi_{_{\rm O}}$ are unperturbed values of the surface density and
gravitational potential, $\Sigma_l$, $\Omega_l$ and $\Phi_l$ are the
perturbation amplitudes, and $\theta$ is an angular coordinate on the surface of
the shell in a direction perpendicular to the plane wave. Substituting these
perturbations (equation \ref{pert}) into the linearised equations
(\ref{hde:ce}--\ref{hde:pe}), and integrating equation (\ref{hde:pe}) over the
thickness of the shell, gives a set of linear differential equations for the
amplitudes $\Sigma_l$ and $\Omega_l$
\begin{equation}\label{linset}
\frac{d}{dt}\left(\begin{array}{c}
\Sigma_l\\
\Omega_l
\end{array}\right)=
\left(\begin{array}{cc}
-\frac{2V}{R}, & -l\Sigma_{_{\rm O}}\\
\frac{lc_{_{\rm S}}^2}{\Sigma_{_{\rm O}}R^2}-\frac{2\pi G}{R}, & -\frac{V}{R}\\
\end{array}\right)
\left(\begin{array}{c}
\Sigma_l\\
\Omega_ll
\end{array}\right)\,.
\end{equation}
The set of equations (\ref{linset}) has two eigenvalues
\begin{equation}\label{omega_thin}
\omega_{_{\rm THIN}}^{(1,2)}(l)=-\frac{3V}{2R}\pm\sqrt{\frac{V^2}{4R^2}+\frac{2\pi G\Sigma_{_{\rm O}}l}{R}-\frac{c_{_{\rm S}}^2l^2}{R^2}}
\end{equation}
and two related eigenvectors
\begin{equation}\label{eigenvectors}
e_l^{(1,2)}=\left(\begin{array}{c}
-l\Sigma_{_{\rm O}}\\
\omega^{(1,2)}+\frac{2V}{R}
\end{array}\right)\,.
\end{equation}
Any harmonic plane-wave perturbation of the shell described by amplitudes
$\Sigma_l$ and $\Omega_l$ can be written in the form
\begin{equation}
\left(\begin{array}{c}\Sigma_l\\\Omega_l\end{array}\right)=Ae_l^{(1)}+Be_l^{(2)}
\ ,
\end{equation}
where $A$ and $B$ are real numbers. Its evolution in the linear regime is then
given by
\begin{equation}
\left(\begin{array}{c}\Sigma\\\Omega\end{array}\right)=Ae_l^{(1)}\exp(\omega^{(1)}t)+Be_l^{(2)}\exp(\omega^{(2)}t)\,.
\end{equation}
If a perturbation is given by the eigenvector (2) only (i.e. A = 0), the mode
is always damped (since $\omega^{(2)}$ is always negative). On the other hand,
if some perturbation is described by eigen-vector (1), it will grow if the
    gravity term in $\omega^{(1)}$ is large enough.

\subsection{Fragment in the thick shell}\label{sec:hore}

It turns out that the thin-shell approximation is a good description of the
linear instability at wavelengths larger than the shell thickness, provided the
shell thickness is uniform. However, in Paper~I we find that the shell thickness
varies with the surface density of the perturbation, even if its amplitude is
relatively small. The spatially varying shell thickness has two effects. (i) In
a low-pressure environment, the gas at the centre of a fragment expands in the
direction perpendicular to the shell surface, and this results in a slower
growth rate and a smaller range of unstable wavelengths, compared with the
predictions of the thin-shell approximation. (ii) In a high-pressure
environment, the external pressure helps to compress fragments from the side,
and this results in a higher growth rate and a larger range of unstable
wavelengths, compared with the predictions of the thin-shell approximation.

In order to take the variable thickness of the shell into account, we model a
fragment as a uniform-density oblate spheroid with major axis $r$ and minor axis
$z$, embedded in an ambient medium with pressure $P_{_{\rm EXT}}$. Radial
excursions of the spheroid are then controlled by the equations of motion
\begin{eqnarray}\nonumber\label{ddr}
\ddot{r}(r,z)&\simeq&-\,\frac{3Gm}{2}\;\left\{\frac{r\,\cos^{-1}(z/r)}{\left(r^2-z^2\right)^{3/2}}\,-\,\frac{z/r}{\left(r^2-z^2\right)}\right\}\\
&&\hspace{2.5cm}-\,\frac{20\pi P_{_{\rm EXT}}rz}{3m}\,+\,\frac{5c_{_{\rm S}}^2}{r}
\end{eqnarray}
and
\begin{eqnarray}\nonumber\label{ddz}
\ddot{z}(r,z)&\simeq&-\,3Gm\;\left\{\frac{1}{\left(r^2-z^2\right)}\,-\,\frac{z\,\cos^{-1}(z/r)}{\left(r^2-z^2\right)^{3/2}}\right\}\\
&&\hspace{2.5cm}-\,\frac{20\pi P_{_{\rm EXT}}r^2}{3m}\,+\,\frac{5c_{_{\rm S}}^2}{z}
\end{eqnarray}
\citep[see][]{2005A&A...430.1059B}.

We assume that the mass of the spheroid, $m$, is constant, and originates from a
comoving circular patch on the shell, with constant angular wavenumber $l$, and
hence radius $\pi R(t)/l$.

The evolution of the spheroid can be followed by solving equations (\ref{ddr})
and (\ref{ddz}) numerically, as in \citet{2005A&A...430.1059B}. However, in
order to derive an analytic dispersion relation for PAGI, we define $t=0$ as the
time when fragmentation starts, and then we assume that for a certain time,
$t_\epsilon$, thereafter the spheroid collapses with constant acceleration
$\ddot{r}\OO$ given by equation (\ref{ddr}) with $r = r\OO$. Here $r\OO=\pi R/l$
is the initial radius of the fragment, and its initial expansion speed is
$\dot{r}\OO =\pi V/l$, where $V$ is the radial velocity of the shell. The
subscript $\epsilon$ denotes the fraction by which the fragment radius shrinks
during $t_\epsilon$ (i.e. $r(t_\epsilon) = (1-\epsilon)r\OO$). The evolution of
the spheroid radius is then given by
\begin{equation}\label{sph_evol}
r(t)=r\OO+\dot{r}\OO t+\frac{1}{2}\ddot{r}\OO t^2\,.
\end{equation}
and hence $t_\epsilon$ is given by
\begin{eqnarray}\label{sph_evol2}
\ddot{r}\OO t_\epsilon^2\,+\,2\dot{r}\OO t_\epsilon\,+\,2\epsilon r\OO&=&0\,.
\end{eqnarray}

In the thin-shell analysis, the instability growth rate is defined as the
inverse of the time it takes for the perturbation described by a given
eigenvector to grow by a factor $e$. Since eigenvectors are not available for
PAGI, we define the instability growth rate simply as
\begin{equation}\label{def_omegaeps}
\omega_\epsilon=\frac{1}{t_\epsilon}\,.
\end{equation}
Since $\omega_\epsilon$ depends on the choice of $\epsilon$, its absolute value
is somewhat arbitrary. However, relative values of $\omega_\epsilon$ for
different wavenumbers are well defined. Moreover, the range of unstable
wavenumbers is independent of $\epsilon$ and can therefore be compared directly
with the range predicted by the thin-shell dispersion relation $\omega_{_{\rm
THIN}}$. Combining Eqns. \ref{def_omegaeps} and  \ref{sph_evol2}, we obtain the
PAGI dispersion relation in the form
\begin{equation}\label{omega_conddr}
\omega_\epsilon=-\frac{\dot{r}\OO}{2\epsilon r\OO}+\sqrt{\left(\frac{\dot{r}\OO}{2\epsilon r\OO}\right)^2-\frac{\ddot{r}\OO}{2\epsilon r\OO}}\,.
\end{equation}

\begin{figure}
\begin{center}
\includegraphics[width=8cm]{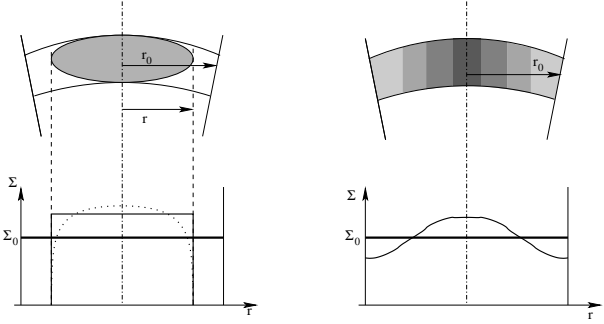}
\end{center}
\caption{Comparison of fragment models in the thin shell approach (right) and
the thick shell approach (left). The approximations are: sine-shaped surface
density and velocity perturbation (thin) vs. uniform oblate spheroid (thick);
and solution of linearised hydrodynamic equations (thin) vs. constant
acceleration coming from non-linear equations of motion (thick).}
\label{fig:shell_models}
\end{figure}

Figure \ref{fig:shell_models} illustrates the models used in the two analyses of
shell gravitational instability. In the thin-shell analysis, the fragment is
modelled as a sine-shaped perturbation of the surface-density and the velocity,
on an infinitesimally thin shell. Its growth rate is obtained by solving
linearised hydrodynamic equations. In the PAGI analysis, the fragment is
modelled as a uniform-density oblate spheroid contained by a constant ambient
pressure. It is assumed to contract with constant acceleration given by the
non-linear equations of motion evaluated at the onset of instability.

\subsection{Dispersion relation of the thick shell}\label{sec:disprel}

The final form of the dispersion relation is obtained by substituting $m\OO =\pi
r\OO^2 \Sigma\OO$, $r\OO =\pi R\OO/l$, $\dot{r}\OO =\pi V\OO/l$ (where
$\Sigma\OO$, $R\OO$ and
$V\OO$ are -- respectively -- the surface-density, radius, and radial velocity of
the shell when the instability commences), and $\ddot{r}\OO = \ddot{r}(r\OO
,z\OO )$ (Eqn. \ref{ddr}) with $z\OO = \Sigma c_{_{\rm S}}^2/\left(2P_{_{\rm
EXT}}+\pi G\Sigma\OO^2\right)$ (Eqn. \ref{zus}). This yields

\begin{eqnarray}\nonumber\label{omega_epsilon}
\omega_\epsilon\!&\!=\!&\!-\frac{V\OO}{2\epsilon R\OO}
+\left\{\left(\frac{V\OO}{2\epsilon R\OO}\right)^2\right.\\\nonumber
&&\hspace{1.35cm}+\,\frac{3\pi G\Sigma\OO}{4\epsilon}\left[\frac{r\OO^2\,\cos^{-1}
\left(z\OO / r\OO\right)}{\left(r\OO^2-z\OO^2\right)^{3/2}}\,
-\,\frac{z\OO}{\left(r\OO^2-z\OO^2\right)}\right]\\\nonumber
&&\hspace{1.35cm}+\frac{10P_{_{\rm EXT}}c_{_{\rm S}}^2 l^2}{3\pi^2\epsilon
R\OO^2(2P_{_{\rm EXT}}+\pi G\Sigma\OO^2)}\\
&&\hspace{1.35cm}\left.-\,\frac{5 c_{_{\rm S}}^2 l^2}{2\pi^2\epsilon R\OO^2}\right\}^{1/2}\,.
\end{eqnarray}

Equation (\ref{omega_epsilon}) has a very similar structure to the thin-shell
dispersion relation (Eqn. \ref{omega_thin}), in that the first two terms on the
right hand side (line one) represent stretching due to expansion, the third term
(line two) represents self-gravity, and the final term (line four) represents
internal pressure. However, there are significant differences. (i) The terms all
depend on the value of $\epsilon$. We revisit this issue below. (ii) The
stretching terms have different numerical coefficients because the initial
conditions for the analyses are different. In the thin-shell analysis, the
initial perturbation consists of harmonic waves in both surface-density and
velocity, given by eigenvectors (Eqn. \ref{eigenvectors}). In the PAGI analysis,
the fragment starts life as a uniform-density spheroid expanding with the shell.
(iii) The self-gravity term reflects the vertical height of the spheroid,
$z\OO$. For a spherical fragment ($z\OO\!=\!r\OO$), the term in the square
bracket is $2/3r\OO$. As the fragment becomes flatter, this term increases, and
in the limit of an infinitesimally thin fragment ($z\OO\!=\!0$) it becomes
$\pi/2r\OO$ (an increase of $3\pi/4\simeq2.36$). (iv) The fourth term in Eqn.
\ref{omega_epsilon} (line three) represents the effect of external pressure, and
therefore has no equivalent in Eqn. \ref{omega_thin}. 

\begin{figure}
\includegraphics[width=8cm]{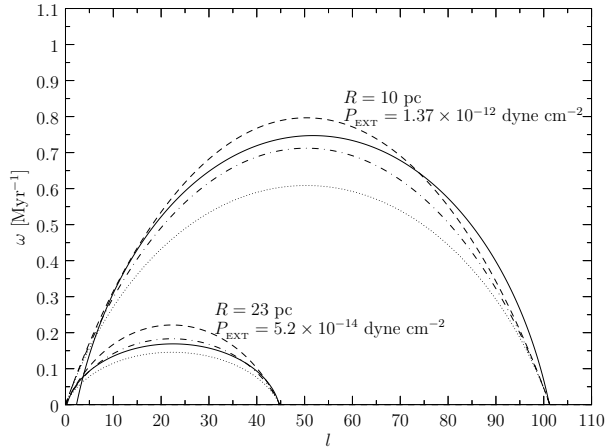}
\caption{Comparison of the thin shell dispersion relation
(Eq.~(\ref{omega_thin}), solid lines) and the PAGI dispersion relation
(Eq.~\ref{omega_epsilon}) for three values of $\epsilon$: $\epsilon = 0.05$
(dashed), $\epsilon = 0.1$ (dash-dotted) and $\epsilon = 0.2$ (dotted). Two
states of the shell with parameters from Paper~I are shown: the initial state
($R = 10$~pc) and the state with the maximum radius $R = 23$~pc. For each state,
the external pressure $P_{_{\rm EXT}} = P_{_{\rm THIN}}$ was calculated so that
$\omega_{_{\rm THIN}}(l)$ and $\omega_\epsilon(l, P_{_{\rm THIN}})$ give the
same range of unstable modes.} 
\label{epsdep} 
\end{figure}

Due to the dependence of the absolute value of the PAGI dispersion relation
on $\epsilon$ it is not possible to compare $\omega_\epsilon$ to $\omega_{_{\rm
THIN}}$ directly (we remind the reader that only the absolute value of $\omega_\epsilon$
depends on $\epsilon$, the range of unstable wavelengths as well as the relative
growth rates of individual modes are $\epsilon$-independent). However, we can
define a critical pressure $P_{_{\rm THIN}}$ such that with $P_{_{\rm
EXT}}=P_{_{\rm THIN}}$ the range of unstable wavenumbers predicted by the PAGI
analysis is the same as that predicted by the thin-shell analysis. Then, for
this choice of $P_{_{\rm EXT}}$, we find $\epsilon$ for which Eqns.
(\ref{omega_epsilon}) and (\ref{omega_thin}) predict very similar growth rates
($\omega_\epsilon(l, P_{_{\rm THIN}})\simeq\omega_{_{\rm THIN}}(l)$). This is
shown by Figure~\ref{epsdep}, for the shell from Paper~I in two states: the
initial radius $R = 10$~pc and the maximum radius $R=23$~pc. It can be seen that
$\omega_\epsilon(l, P_{_{\rm THIN}})\simeq\omega_{_{\rm THIN}}(l)$ if we set
$\epsilon = 0.1$, for both shell radii. Moreover, this value of $\epsilon$
results in both dispersion relations being close to each other throughout the
whole shell expansion and we therefore adopt this as our default value of
$\epsilon$.

\begin{figure*}
\includegraphics[width=8cm]{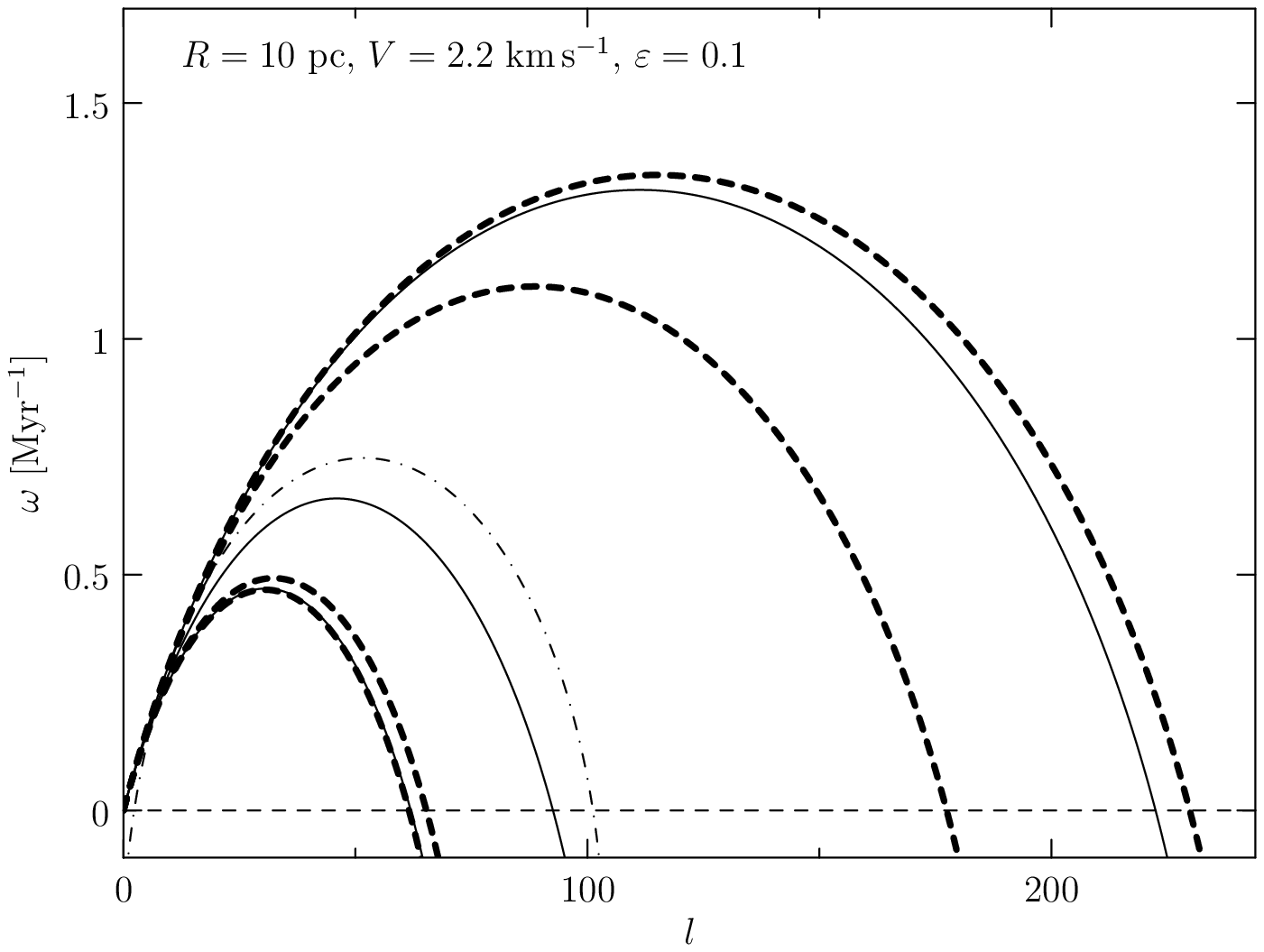}
\includegraphics[width=8cm]{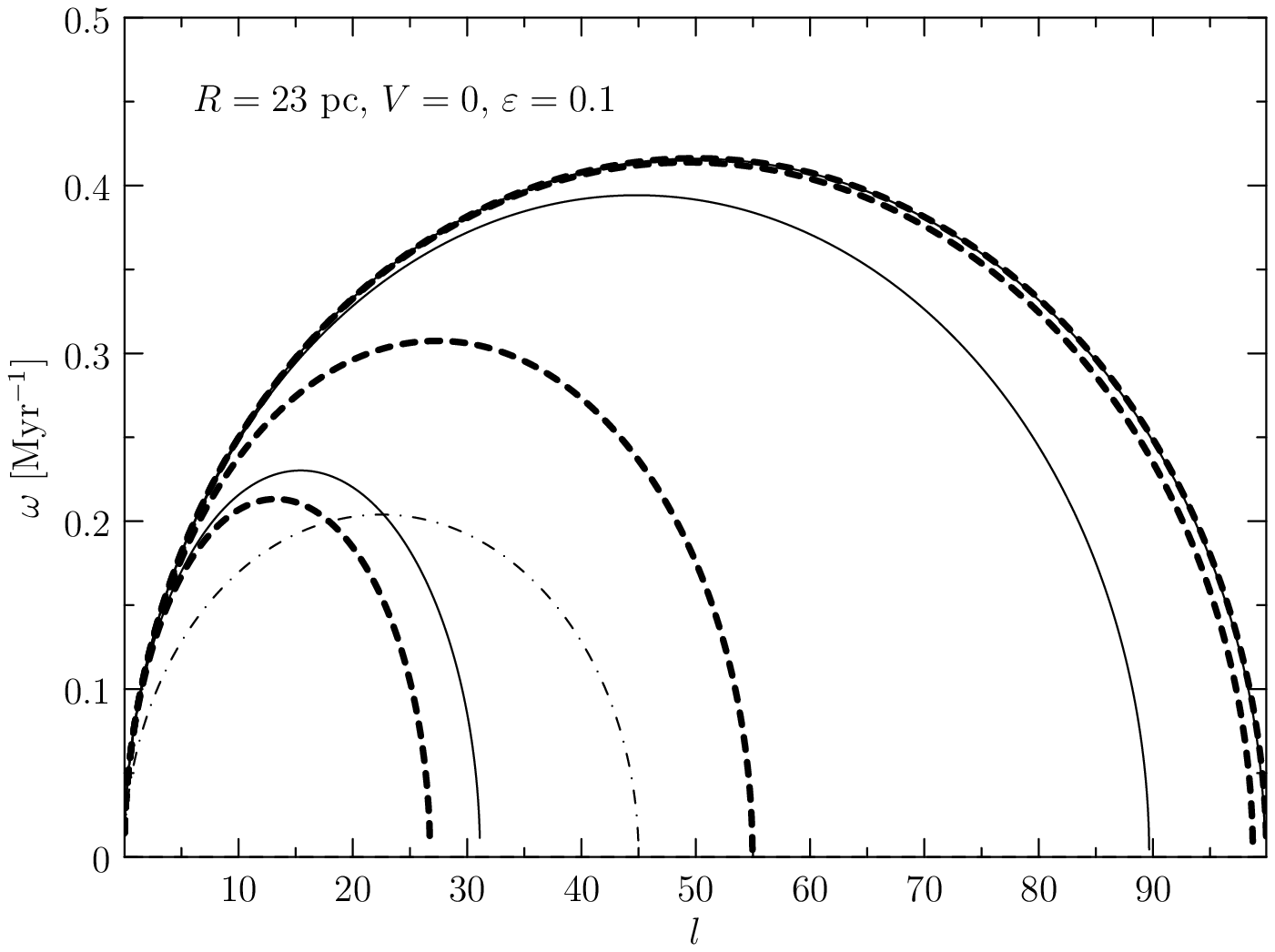}
\caption{Perturbation growth rate for different values of external pressure for
two states of the shell from paper I: $R = 10$~pc, $V = 2.2$~km\,s$^{-1}$
(left), $R = 23$~pc, $V = 0$ (right). Alternating dashed and solid lines denote
$\omega_\epsilon$ for $P_{_{\rm EXT}} = 0$, $10^{-14}$, $10^{-13}$, $10^{-12}$,
$10^{-11}$, $10^{-10}$ and $1$~dyne\,cm$^{-2}$ (in this order for growing range
of unstable wavenumbers). The dash-dotted line shows
perturbation growth rate of the thin shell $\omega_{_{\rm THIN}}$. The
horizontal dashed line at $\omega = 0$ represents the separation between
unstable and damping modes.}
\label{omega_fig}
\end{figure*}

Figure ~\ref{omega_fig} compares the PAGI dispersion relation for different
values of $P_{_{\rm EXT}}$ with the thin-shell dispersion relation. We have used
parameters for a shell having a mass of $2\times 10^4\,{\rm M}_\odot$, (a) with
radius $10\,{\rm pc}$ and outward radial velocity $2.2\,{\rm km}\,{\rm s}^{-1}$,
and (b) with radius $23\,{\rm pc}$ and zero outward radial velocity; this is the
same as the shell that we simulated in Paper~I. Increasing $P_{_{\rm EXT}}$
results in a larger range of unstable wavenumbers, {\it and} faster growth
rates, but even as $P_{_{\rm EXT}}\!\rightarrow\!\infty$, the maximum unstable
wavenumber, $l_{_{\rm PAGI}}^{^{\rm MAX}}$ remains finite. Specifically, for
$P_{_{\rm EXT}}=0$, $l_{_{\rm PAGI}}^{^{\rm MAX}}=0.595l_{_{\rm THIN}}^{^{\rm
MAX}}$, and for $P_{_{\rm EXT}}=\infty$, $l_{_{\rm PAGI}}^{^{\rm
MAX}}=2.2l_{_{\rm THIN}}^{^{\rm MAX}}$, where $l_{_{\rm THIN}}^{^{\rm MAX}}$ is
the maximum unstable wavenumber according to the thin-shell dispersion relation;
evidently an infinitesimally thin shell (i.e. PAGI with $P_{_{\rm EXT}}=\infty$)
does not give the same dispersion relation as the thin-shell model.

\begin{figure}
\begin{center}
\includegraphics[width=8cm]{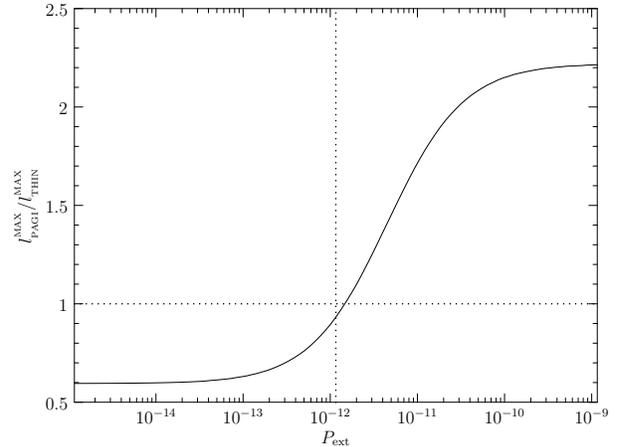}
\end{center}
\caption{Ratio of the shortest unstable modes given by the thick shell and thin
shell models as a function of the external pressure $P_{_{\rm EXT}}$. The
vertical dotted line denotes value of $P_{_{\rm CRIT}}$ for which the
shell has the maximum thickness for a given surface density.}
\label{l0rat}
\end{figure}

Figure~\ref{l0rat} shows how the ratio $l_{_{\rm PAGI}}^{^{\rm MAX}}/l_{_{\rm
THIN}}^{^{\rm MAX}}$ varies with $P_{_{\rm EXT}}$. The vertical line denotes the
critical external pressure, $P_{_{\rm CRIT}}=\pi G\Sigma\OO^2/2$, at which the
derivative of the shell thickness $\left(z_{_{\rm US}}\right)$ with respect to
its surface density $\left(\Sigma_{_{\rm 0}}\right)$ is zero, i.e. the
configuration in which external pressure and self gravity make equal
contributions to containment of the shell, and the shell thickness is only
weakly dependent on its surface-density. This is why $P_{_{\rm CRIT}}$ is
approximately equal to $P_{_{\rm THIN}}$ (where $l_{_{\rm PAGI}}^{^{\rm
MAX}}/l_{_{\rm THIN}}^{^{\rm MAX}}=1$). With this external pressure, as a
fragment starts to form, and its surface-density increases, its thickness does
not change much, and therefore the external pressure force remains largely
perpendicular to the surface of the shell and does not make a significant
contribution to the lateral squashing of the fragment. 

\section{Comparison with numerical simulations}\label{sec:num}

In Paper~I we have studied numerically the evolution of a shell with mass
$M=2\times 10^4\,{\rm M}_\odot$, temperature $T=10\,{\rm K}$, initial radius
$R_\mathrm{ini}=10\,{\rm pc}$ and initial expansion velocity
$V_\mathrm{ini}=2.2\,{\rm km}\,{\rm s}^{-1}$, embedded in a rarefied medium with
pressure $P_{_{\rm EXT}}$. After about $18\,{\rm Myr}$, the expansion stalls, at
radius $R_\mathrm{max}=23\,{\rm pc}$, and the shell starts to collapse. 

We have run simulations with three different external pressures $P_{_{\rm
EXT}}=10^{-17}$, $P_{_{\rm EXT}}=10^{-13}$ and $P_{_{\rm EXT}}=5\times
10^{-13}\,{\rm dyne}\,{\rm cm}^{-2}$; the first one is as close to zero as
the hydrodynamic solver permits, the second one is chosen so that the
shell thickness varies as little as possible during the shell evolution, 
and the third one is the highest pressure for which the shell thickness can be
resolved with the available computational power. 

\begin{figure*}
\includegraphics[width=5.5cm]{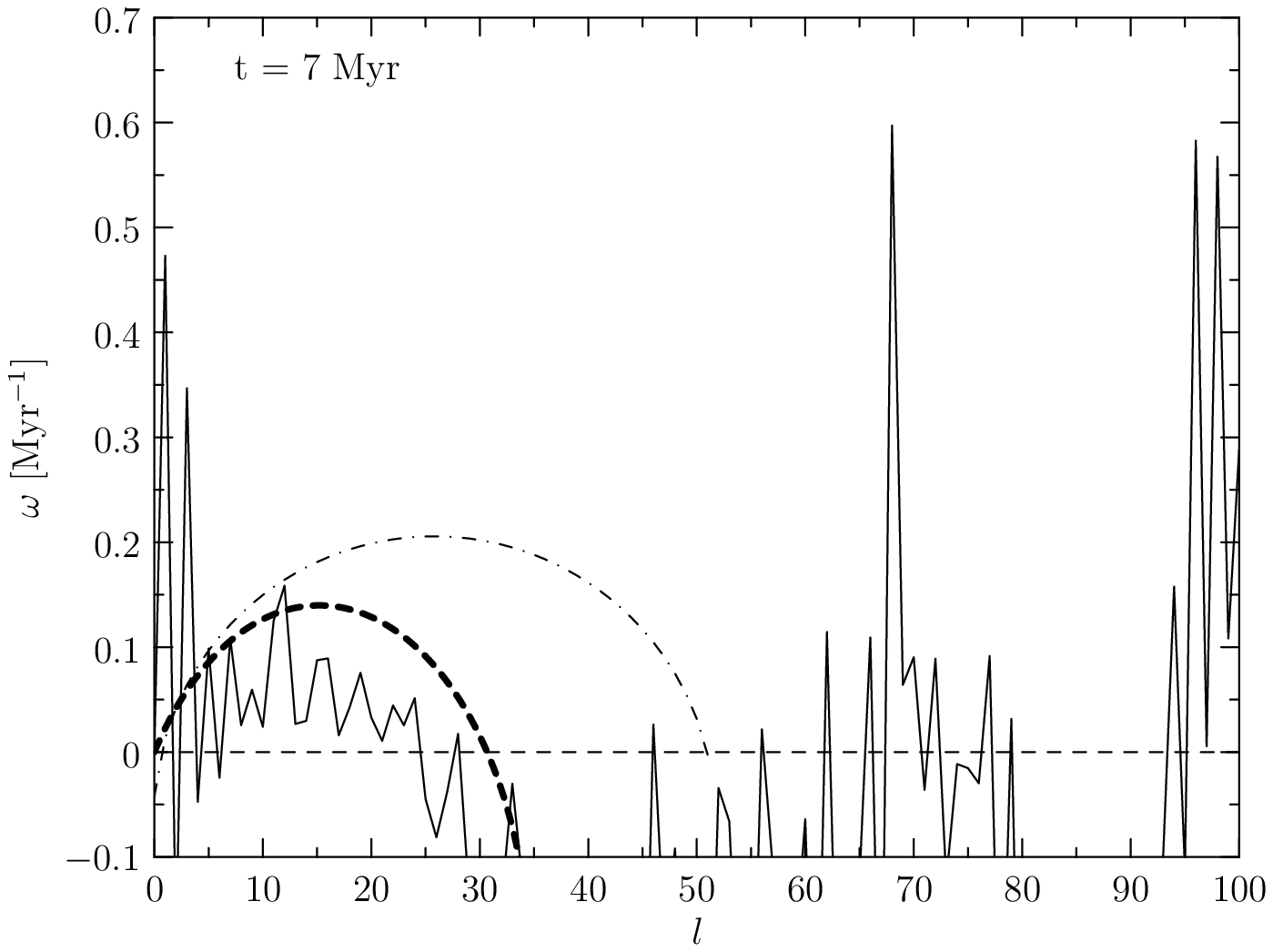}
\includegraphics[width=5.5cm]{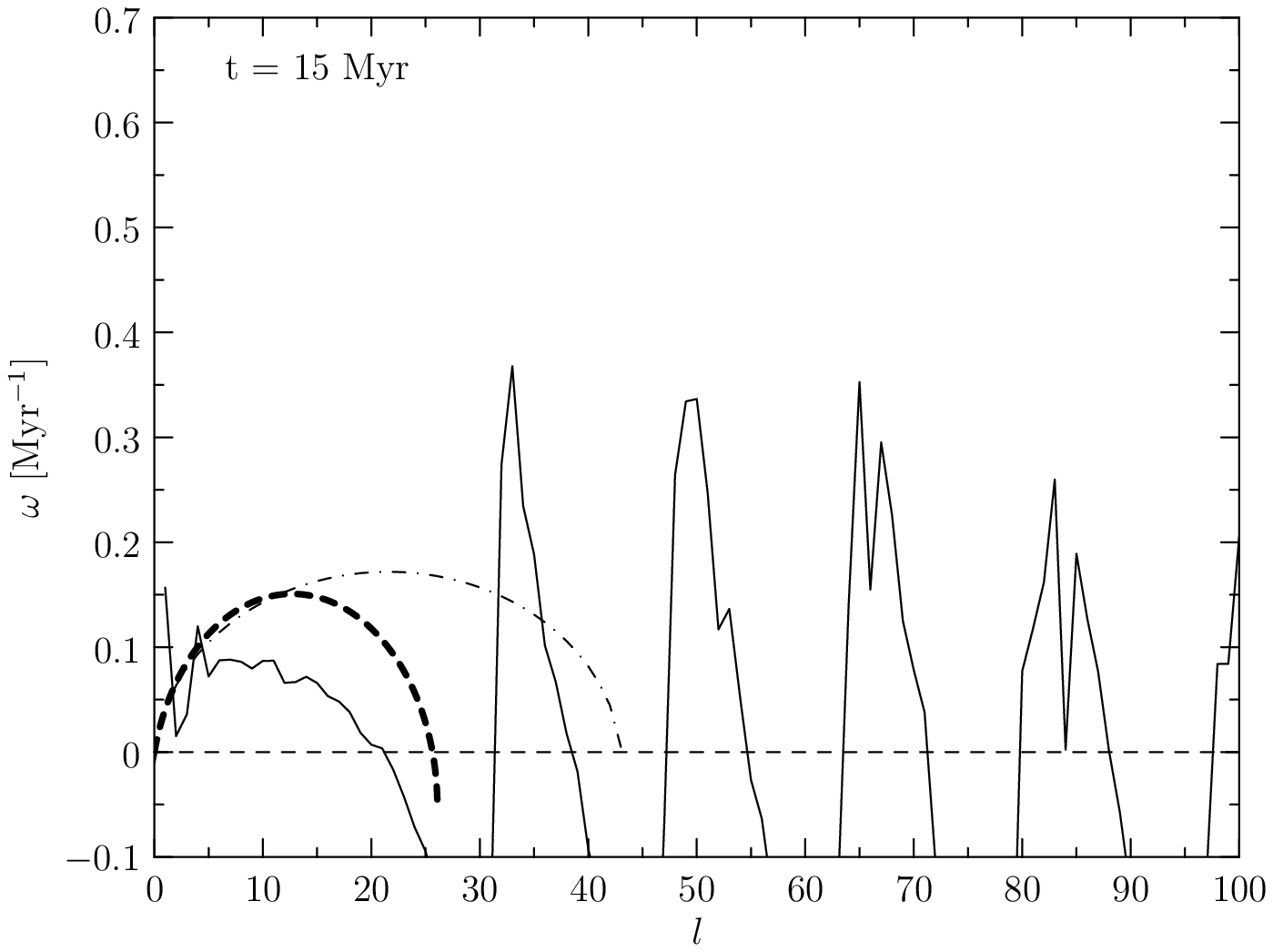}
\includegraphics[width=5.5cm]{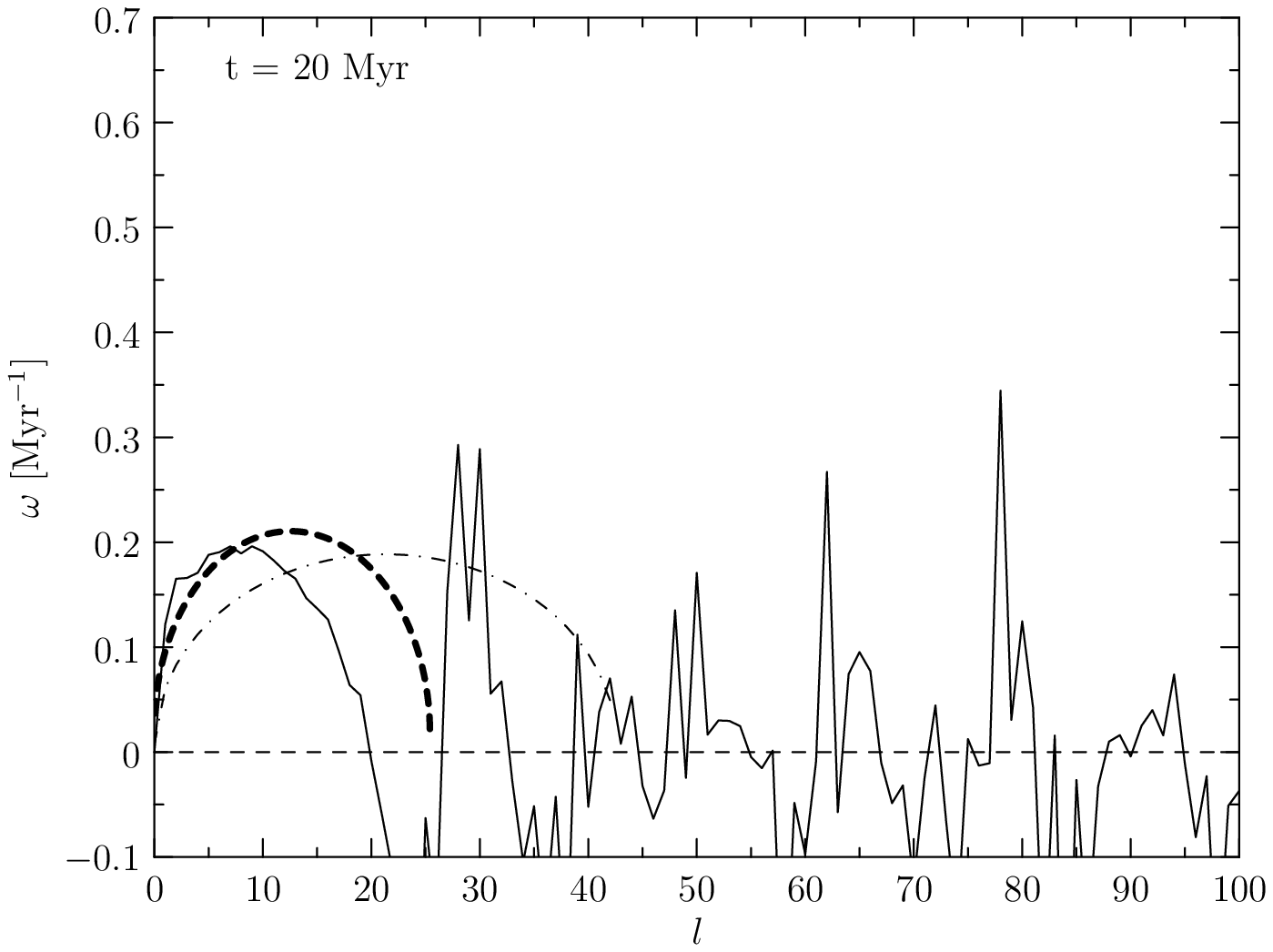}
\includegraphics[width=5.5cm]{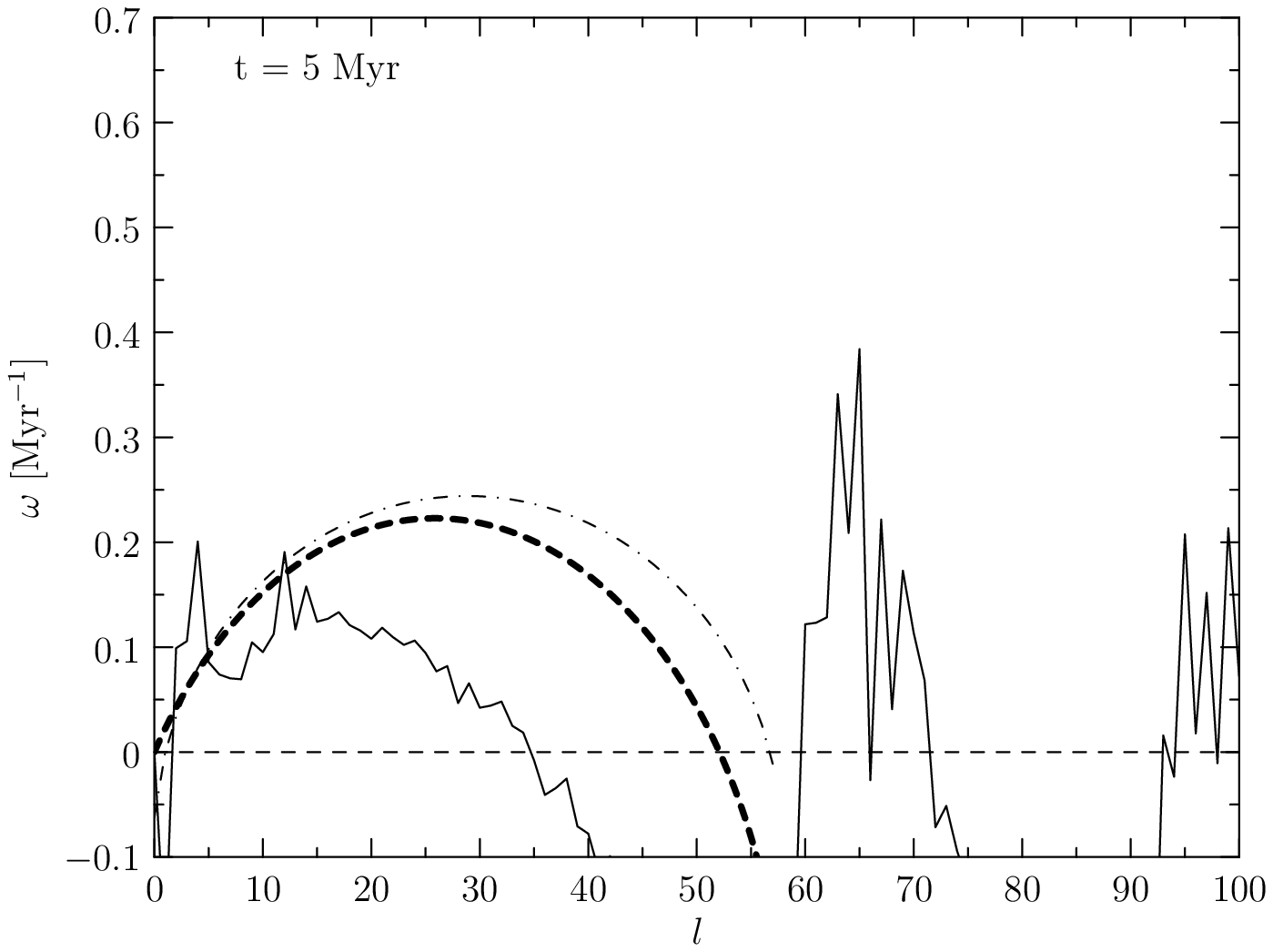}
\includegraphics[width=5.5cm]{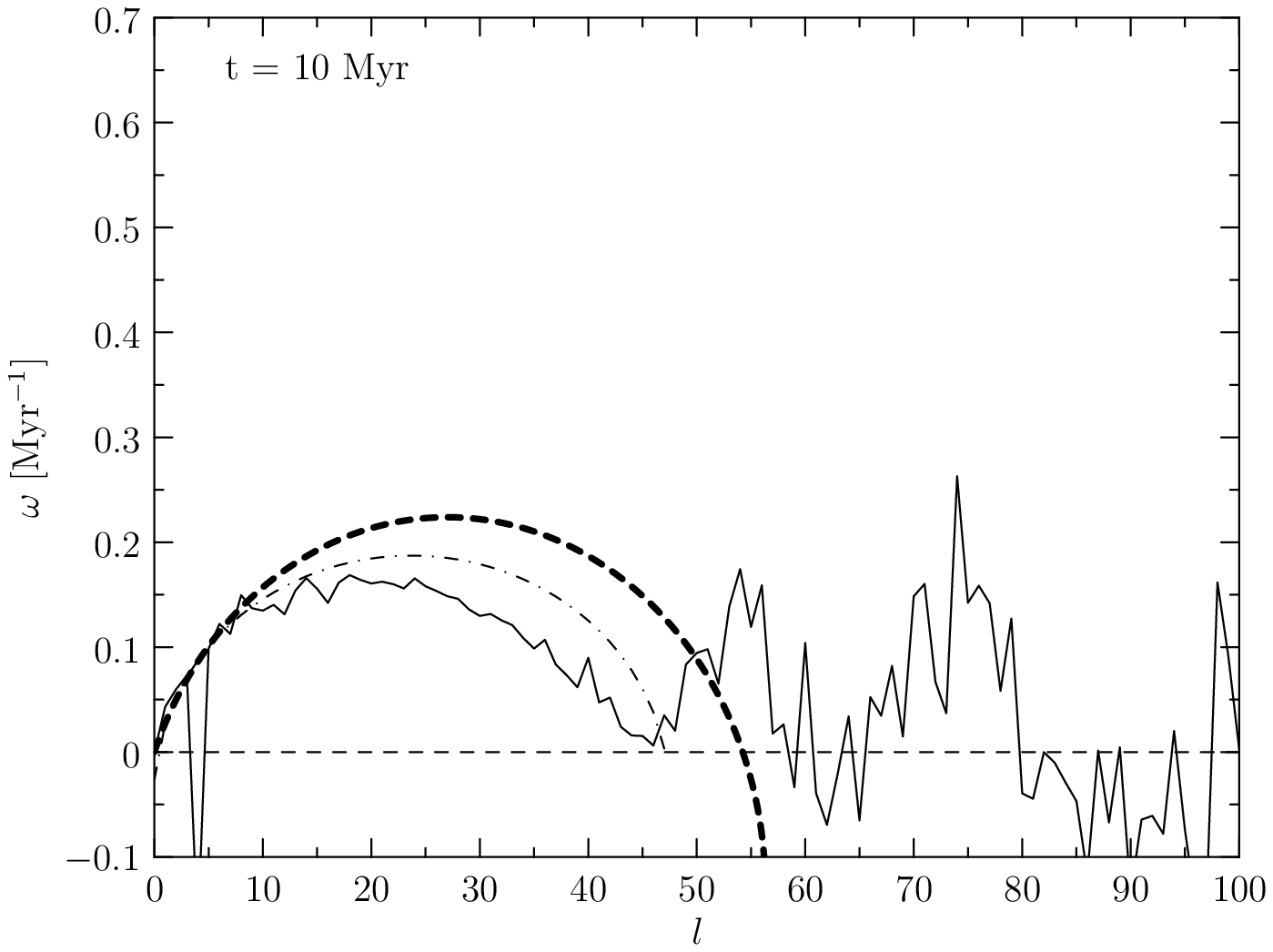}
\includegraphics[width=5.5cm]{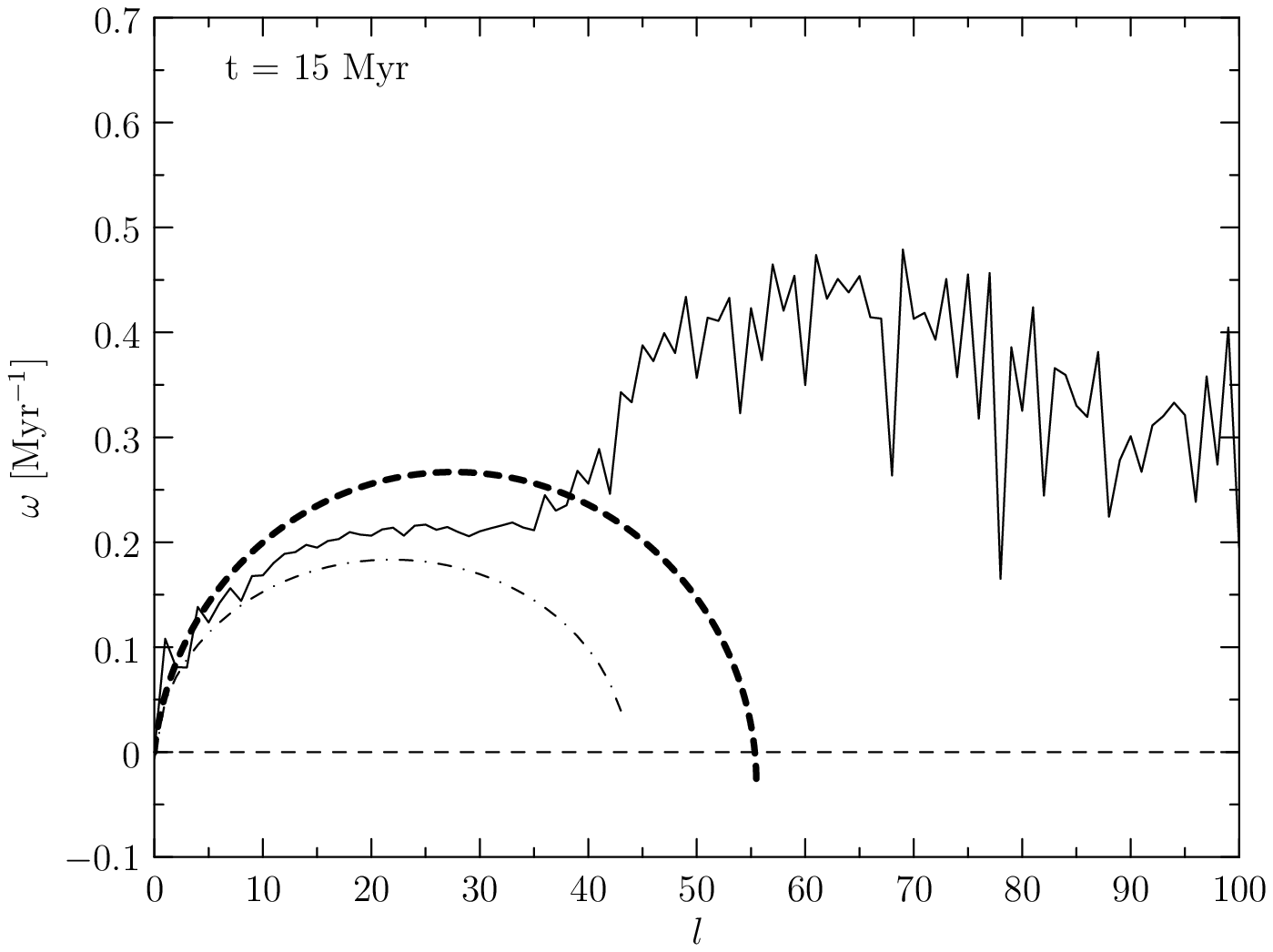}
\includegraphics[width=5.5cm]{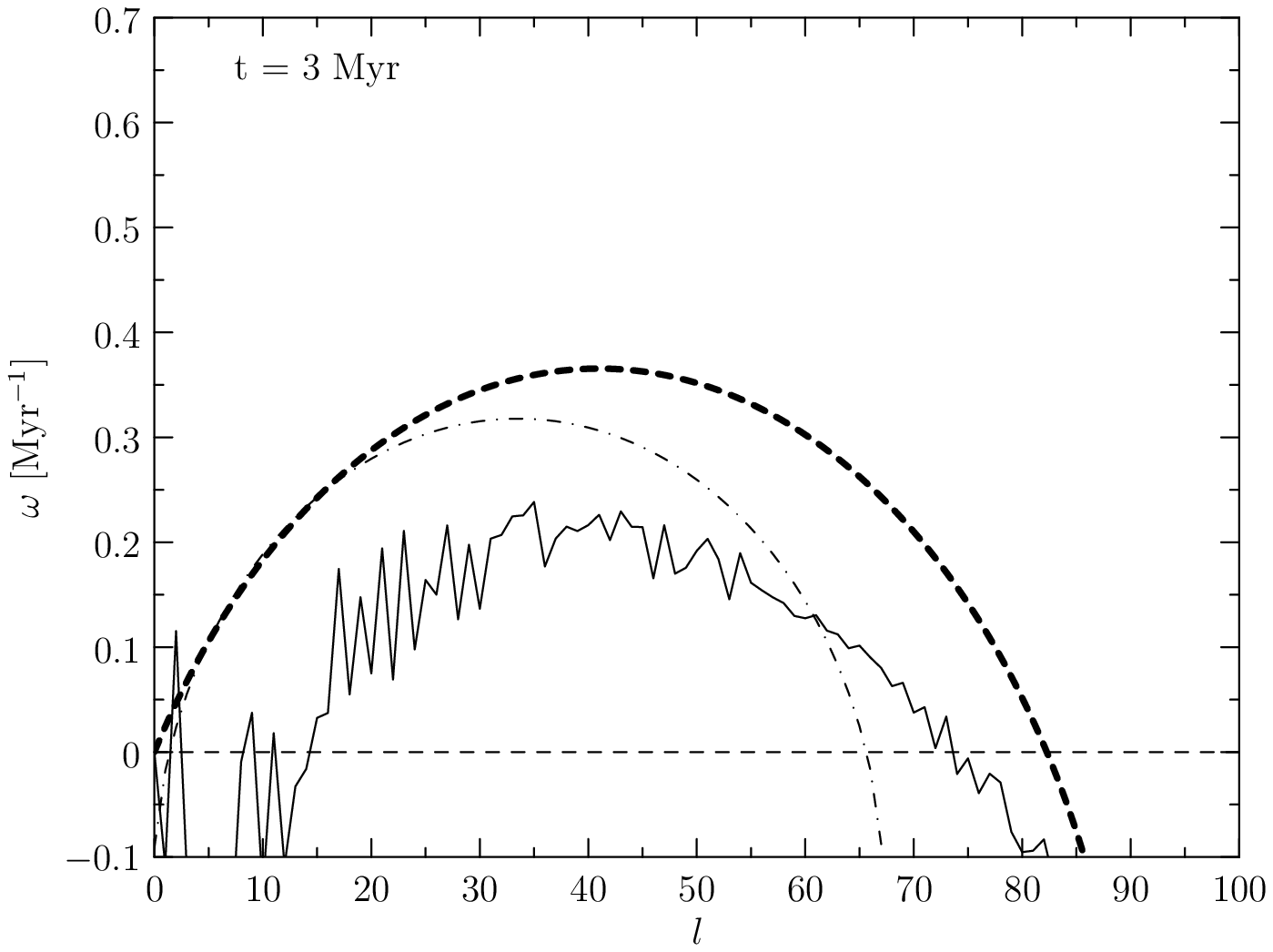}
\includegraphics[width=5.5cm]{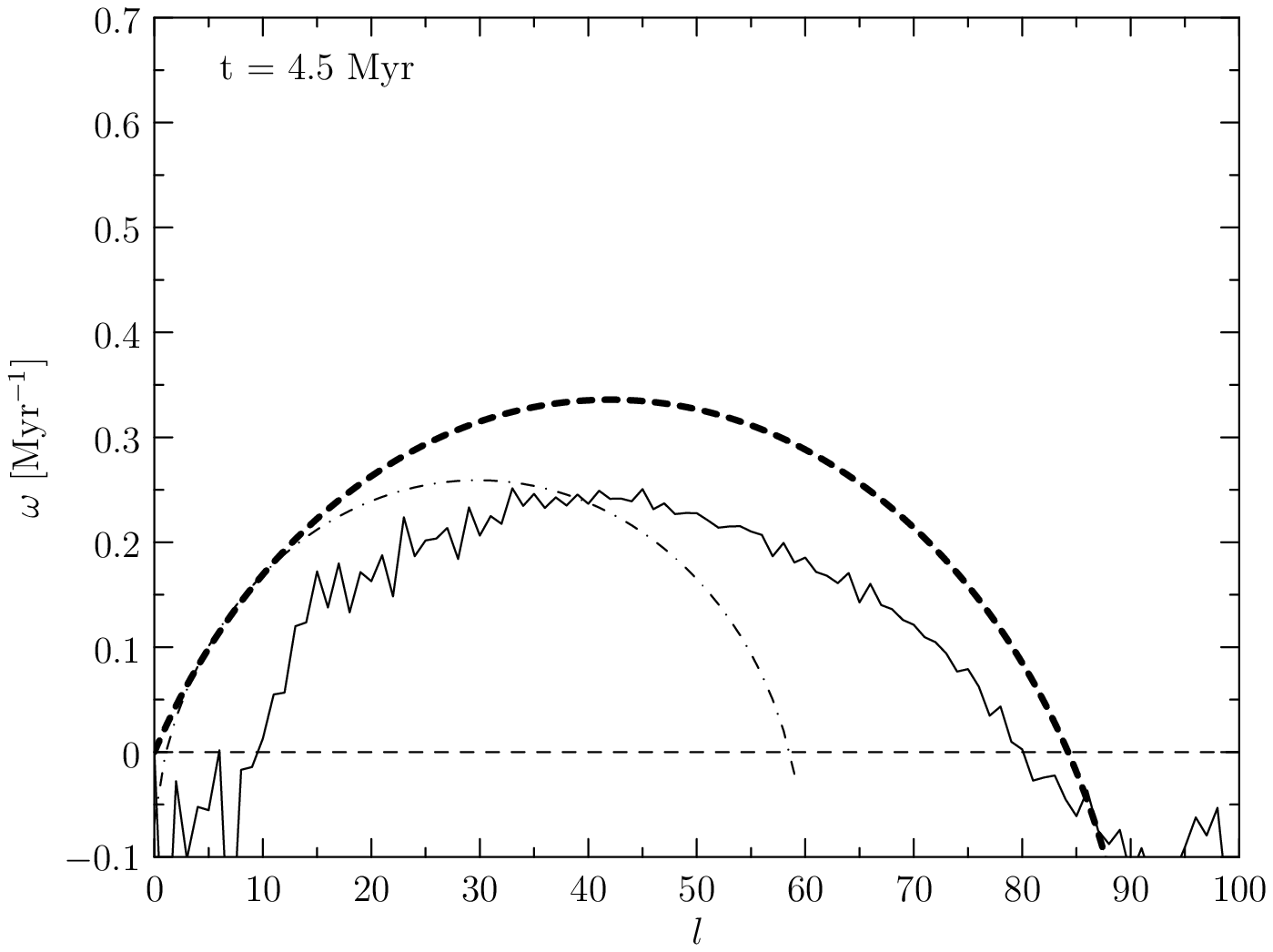}
\includegraphics[width=5.5cm]{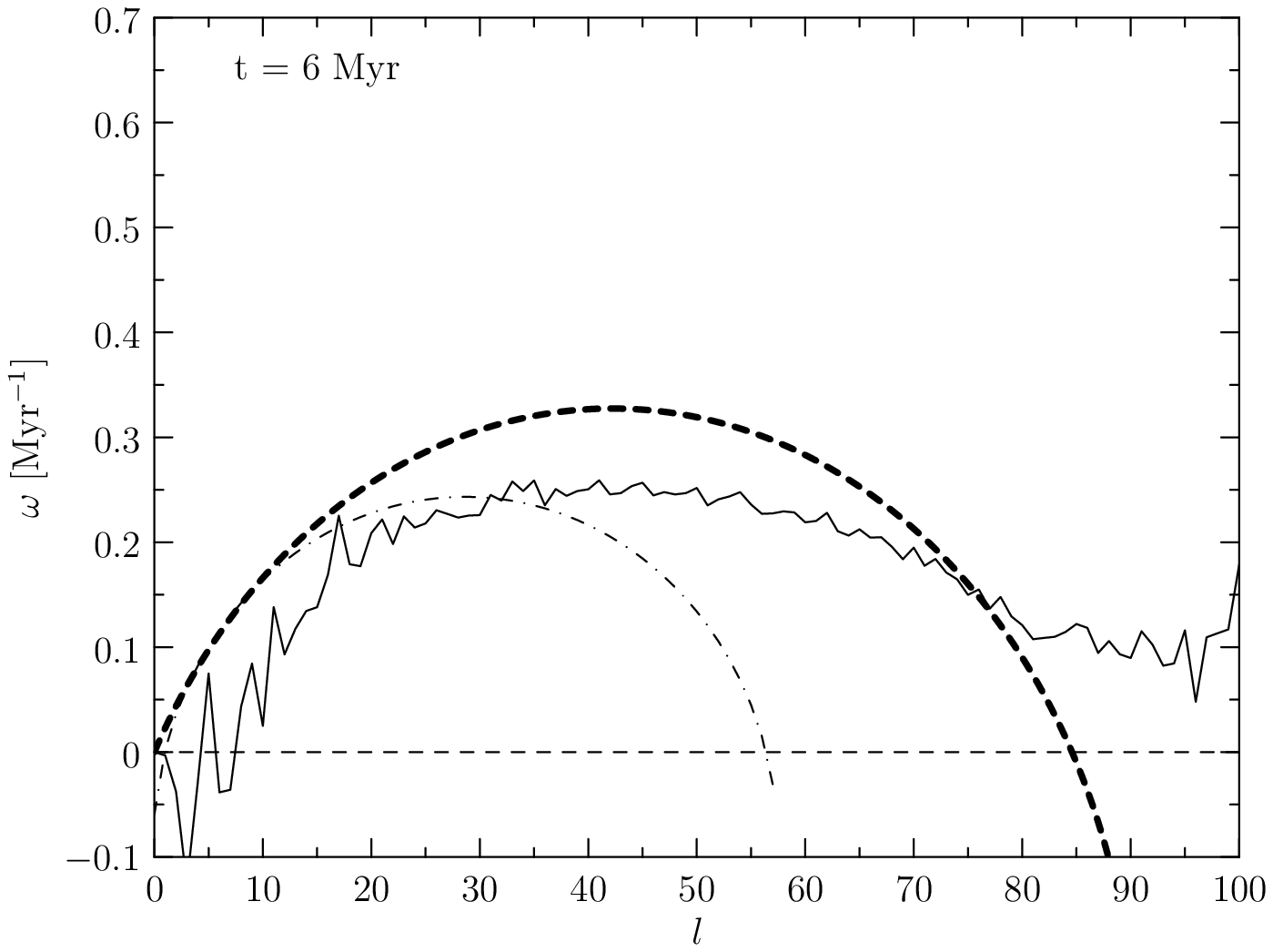}
\caption{The perturbation growth rate in hydrodynamic simulations, compared to both the
thin-shell and the PAGI dispersion relations. The top panels show the simulation
with $P_{_{\rm EXT}}=10^{-17}\,{\rm dyne}\,{\rm cm}^{-2}$; the middle panels
show the simulation with $P_{_{\rm EXT}}=10^{-13}\,{\rm dyne}\,{\rm cm}^{-2}$;
and the bottom panels show the simulation with $P_{_{\rm EXT}}=5\times
10^{-13}\,{\rm dyne}\,{\rm cm}^{-2}$. In all panels only the first
(approximately parabolic) peak at low wavenumbers should be compared with the
analysis. The other peaks are higher harmonics, which reflect the shapes adopted
by fragments in the non-linear condensation regime. The
horizontal dashed line at $\omega = 0$ represents the separation between
unstable and damping modes.
}
\label{omega_cmp}
\end{figure*}

\begin{figure*}
\includegraphics[width=5.5cm]{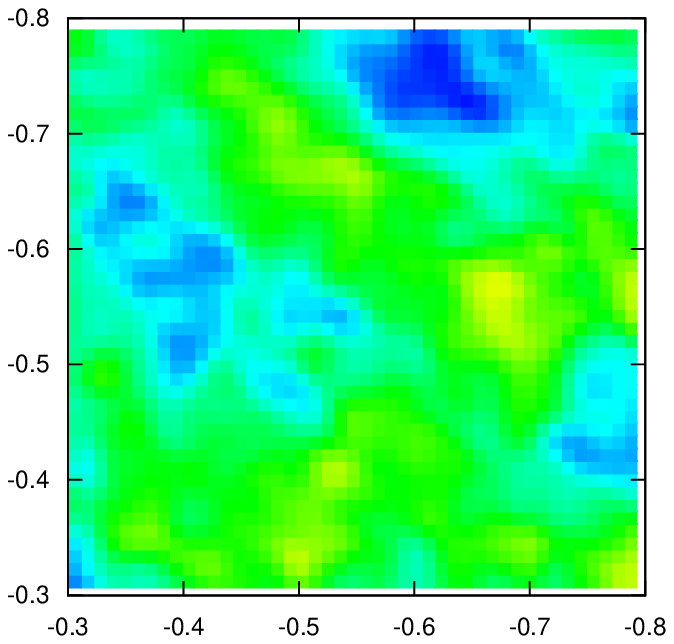}
\includegraphics[width=5.5cm]{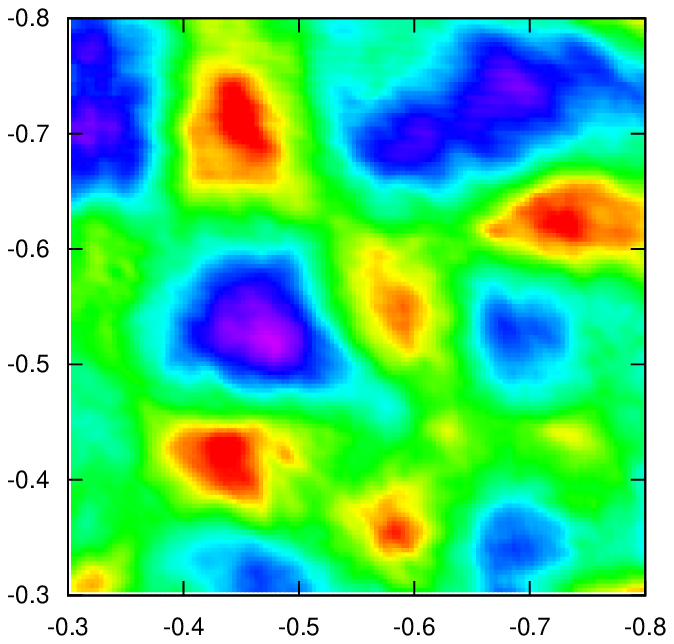}
\includegraphics[width=5.5cm]{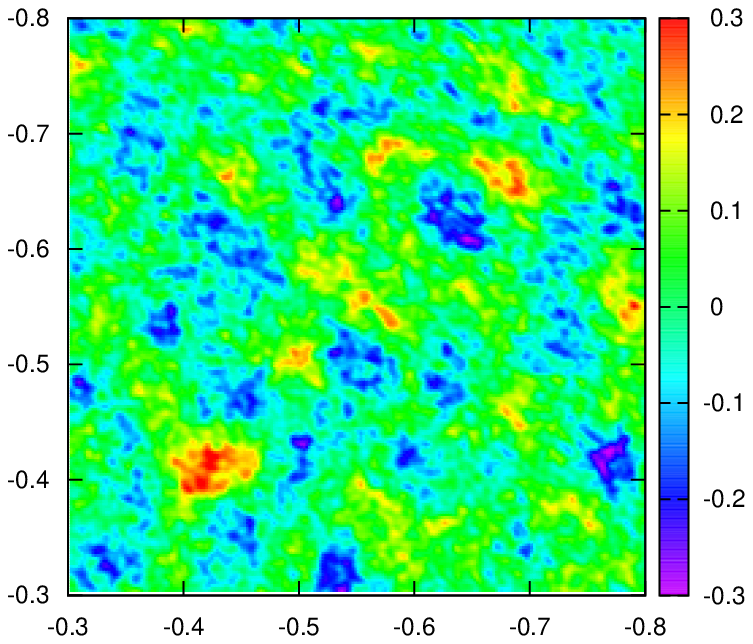}
\caption{Normalised surface density ($\Sigma$/$\Sigma_{_{\rm O}}$) for the three
hydrodynamic simulations with different external pressures:
$P_{_{\rm EXT}} = 10^{-17}$ (left), $P_{_{\rm EXT}} = 10^{-13}$ (middle) and
$P_{_{\rm EXT}} = 5\times 10^{-13}$~dyne\,cm$^{-2}$ (right). Panels show a randomly chosen
part of the shell surface of the size $0.4\times 0.4$ radians (the axes are
denoted in Hammer projection coordinates). The times are the same as for the
middle column of the figure~\ref{omega_cmp}, i.e. $15$, $10$ and $4.5$~Myr for
the low, medium and high pressure simulation, respectively.
}
\label{sigma_cmp}
\end{figure*}

Figures \ref{omega_cmp} and \ref{sigma_cmp} summarize results from the
hydrodynamic simulations. In the low-pressure run, we find that the
perturbation growth rates diverge significantly from the predictions of the
thin-shell dispersion relation $\omega_{_{\rm THIN}}(l)${ -- only the low
wavenumbers grow.} The agreement is better for the medium-pressure
simulation. The high pressure run again diverges from the predictions given by
$\omega_{_{\rm THIN}}(l)$ -- the range of unstable modes includes higher
wavenumbers than predicted. Differences in relative growth rates
among the three simulations result in different fragment sizes as can be seen in
figure~\ref{sigma_cmp}.

\begin{figure}
\begin{center}
\includegraphics[width=8cm]{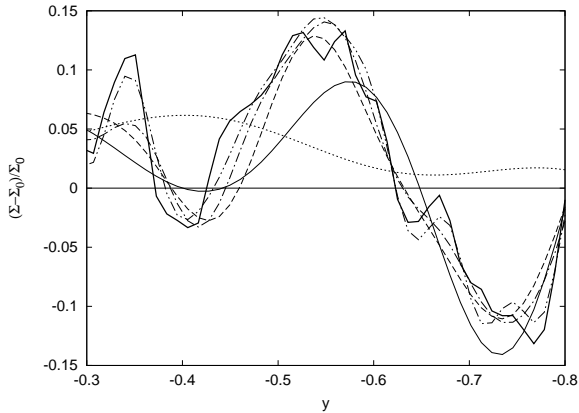}
\end{center}
\caption{Surface density profile of the fragment from the simulation with 
$P_{_{\rm EXT}} = 10^{-17}$~dyne\,cm$^{-2}$ at $t = 15$~Myr (corresponding to the
top middle panel of Figure~\ref{omega_cmp} and the left panel of
Figure~\ref{sigma_cmp}). The profile is taken along the Hammer projection
coordinate at $x = -0.7$. The thick solid line is the actual surface density in
the simulation, the other lines are a sequance of its spherical harmonics
representation with modes up to a certain wavenumber: $l < 15$ (dotted), $l <
30$ (thin solid), $l < 45$ (dashed), $l < 60$ (dash-dotted) and $l < 95$
(dash-doubledotted).}
\label{frag1d}
\end{figure}

The dispersion relation obtained from numerical simulations exhibits
several peaks in some cases (see the top middle panel of
Figure~\ref{omega_cmp}). Among them, only the first peak (at lowest wavenumbers)
is suitable for analysis of mode growth rates. The other peaks represent higher
harmonics of the first peak and their growth is induced by the growth of modes
from the first peak. The higher harmonic peaks describe fragment shapes which
are typically more concentrated than pure spherical harmonics with wavelengths
corresponding to fragment sizes. This is illustrated by Figure~\ref{frag1d}
which shows the surface density profile across an arbitrary fragment in the
simulation with $P_{_{\rm EXT}} = 10^{-17}$~dyne\,cm$^{-2}$ and a sequence of
its representation by spherical harmonics which take into account only modes up
to a certain wavenumber. It can be seen that the modes from the first peak only
($l < 30$) are enough to obtain approximately the correct fragment wavelength.
If the second peak ($l < 45$) is taken into account the representation of the
fragment is already very good. Taking into account modes with $l > 45$ results
in further small improvements in representing the detailed shape of the
fragment.

The perturbation growth rates from the numerical simulations agree much more
closely with the predictions of the PAGI dispersion relation (Eqn.
\ref{omega_epsilon}). The agreement is best at intermediate times. At earlier
times the numerical simulation has not yet had sufficient time to relax and so
the modes are poorly defined. At later times the gravitational instability has
become strongly non-linear. However, the PAGI dispersion relation does tend to
give a slightly larger range of unstable wavenumbers than the numerical
simulations.


\section{Discussion}\label{discussion} 

The PAGI dispersion relation predicts that the fragment mass function depends on
the external pressure confining the shell. Shells expanding into a low pressure
environment, for instance shells at higher galactic latitudes, should fragment
preferentially on long wavelengths. This may result in a top-heavy core
mass function (CMF). However, it remains an open question whether the massive
fragments will form massive stars or whether they will continue to fragment into
smaller pieces. In the low pressure simulation presented in this work, the large
fragments include relatively dense cores (whose growth is represented by
distinctive higher harmonics in measured mode growth rates), and the evolution
exhibits rather merging of fragment cores than fragment splitting. However, this
can be due to our physical model being too simple (isothermal equation of state,
relatively small perturbations of initial density and velocity, no radiation,
stellar feedback, etc.). To answer this question properly, more realistic
simulations would be necessary.

If a top-heavy CMF results in a generation of stars with a top-heavy IMF, this
should be favourable to the production of further shells, and hence to
sequential self-propagating star formation \citep{1977ApJ...214..725E}. 
However, it is a question if such secondary shells expanding into low density
gas are able to collect enough mass, cool down and fragment. An example of such
a shell in a low density environment might be the Carina Flare super-shell
GSH~287+04-17, which extends up to $\sim 450\,{\rm pc}$ above the galactic plane
\citep{1999PASJ...51..751F, 2008PASJ...60.1297D}.

In Paper~I and in this paper, we have studied shells expanding into a hot
rarefied gas. Such shells are confined only by their self gravity and by
external thermal pressure. This model is applicable to shells which break out of
a molecular cloud and expand into a rarefied warm intercloud medium. Our reason
for focusing on this model is that we want to study pure gravitational
instability. By reducing the density of the ambient gas and hence making its
ram pressure negligible, we suppress the typically much faster Vishniac
instability.

For a shell expanding into the relatively dense gas of a molecular cloud, ram
pressure plays an important role in compressing the outer surface of the shell.
The Vishniac instability is then inevitable, because ram pressure acts purely
parallel to the direction of shell expansion, whereas thermal pressure acts
perpendicular to the surface of the shell. Furthermore, ram pressure cannot
compress a fragment laterally, and we therefore expect that in this situation
the growth of small fragments would be slower than predicted by the PAGI
dispersion relation (Eqn. \ref{omega_epsilon}).

The PAGI dispersion relation predicts a range of unstable wavenumbers
which is systematically larger by $\sim 25\%$ than the range obtained from
numerical simulations. This discrepancy appears at high wavenumbers, so it is
unlikely to be due to the curvature of the shell (which is not taken into
account in the derivation of $\omega_\epsilon$), because the high wavenumbers
are least affected by it. The most probable source of this discrepancy is the
approximation of spheroid evolution by collapse (or expansion) with 
constant acceleration (Eq.~\ref{sph_evol}). This is certainly not true, as shown by
\citet{2005A&A...430.1059B}, who found complex behaviour for spheroids in the
non-linear regime. However, the aim of this work is not to give an accurate
description of the fragment evolution, but to obtain an analytical formula which
quickly provides information about the instability of a given wavelength on
the shell and which takes into account the effects of external pressure.


\section{Conclusions}\label{conclusions} 

We have studied an isothermal ballistic shell, confined from both sides by a hot
highly rarefied gas having non-zero pressure, $P_{_{\rm EXT}}$. We have solved
the equation of hydrostatic equilibrium in a frame of reference moving with the
shell and shown that the resulting shell density profiles are in very good
agreement with the profiles obtained from three-dimensional hydrodynamic
simulations. We have also shown that the shell thickness measured at half of its
peak density agrees well with the thickness of a simple uniform-density shell
model.

This allows us to model a fragment forming in the shell, due to gravitational
instability, as a uniform-density oblate spheroid. We use this model to derive a
dispersion relation for pressure assisted gravitational instability (PAGI). This
dispersion relation (perturbation growth rate as a function of initial fragment
wavenumber) depends on the pressure of the external medium: the higher $P_{_{\rm
EXT}}$, the larger the maximum unstable wavenumber (i.e. the smaller the
smallest unstable fragment), and the faster the growth rate for an unstable
fragment. If $P_{_{\rm EXT}}\!=\!0$, the highest unstable wavenumber is $0.6$
times smaller than predicted by the standard thin-shell analysis using the same
parameters; if $P_{_{\rm EXT}}\!=\!\infty$, the highest unstable wavenumber is
$2.2$ times larger. The PAGI dispersion relation gives approximately the same
range of unstable wavelengths as the thin-shell dispersion relation if $P_{_{\rm
EXT}}\simeq P_{_{\rm CRIT}}$, where $P_{_{\rm CRIT}}$ is the critical external
pressure for which self-gravity and external pressure contribute equally to
confinement of the shell.

Finally, we have demonstrated that the predictions of the PAGI dispersion
relation agree rather well with the results of three-dimensional hydrodynamic
simulations. In particular, (i) the PAGI dispersion relation predicts a maximum
unstable wavenumber very similar to the simulations (but systematically $\sim
25\%$ higher); (ii) modulo this offset, the increase in $l_{_{\rm PAGI}}^{^{\rm
MAX}}$ with $P_{_{\rm EXT}}$ predicted by the PAGI dispersion relation is
exactly mirrored by the simulations.


\section{Acknowledgments} 

The authors thank the anonymous referee for his/her very thorough reading of the
paper and constructive comments.
The FLASH code was developed in part by the DOE-supported Alliances Center for
Astrophysical Thermonuclear Flashes at the University of Chicago. The AMR
calculations were performed on the Cardiff University HPC Cluster MERLIN. The
SPH simulations were performed on the Virgo cluster of the Astronomical
Institute of the Academy of Science of the Czech Republic, v. v. i. RW
acknowledges support by the Human Resources and Mobility Programme of the
European Community under contract MEIF-CT-2006-039802. JED acknowledges support
from a Marie Curie fellowship as part of the European Commission FP6 Research
Training Network `CONSTELLATION' under contract MRTN--CT--2006--035890. JED, RW
and JP acknowledge support from the Institutional Research Plan AV0Z10030501 of
the Academy of Sciences of the Czech Republic and project LC06014--Centre for
Theoretical Astrophysics of the Ministry of Education, Youth and Sports of the
Czech Republic. APW acknowledges the support of STFC grant PP/E000967/1.


\bibliography{myrefs}

\begin{thebibliography}{}

\bibitem[\protect\citeauthoryear{{Boyd} \& {Whitworth}}{{Boyd} \&
  {Whitworth}}{2005}]{2005A&A...430.1059B}
{Boyd} D.~F.~A.,  {Whitworth} A.~P.,  2005, \aap, 430, 1059

\bibitem[\protect\citeauthoryear{{Churchwell}, {Povich}, {Allen}, {Taylor},
  {Meade}, {Babler}, {Indebetouw}, {Watson}, {Whitney}, {Wolfire}, {Bania},
  {Benjamin}, {Clemens}, {Cohen}, {Cyganowski}, {Jackson}, {Kobulnicky} \&
  {Mathis}}{{Churchwell} et~al.}{2006}]{2006ApJ...649..759C}
{Churchwell} E.,  {Povich} M.~S.,  {Allen} D.,  {Taylor} M.~G.,  {Meade} M.~R.,
   {Babler} B.~L.,  {Indebetouw} R.,  {Watson} C.,  {Whitney} B.~A.,  {Wolfire}
  M.~G.,  {Bania} T.~M.,  {Benjamin} R.~A.,  {Clemens} D.~P.,  {Cohen} M.,
  {Cyganowski} C.~J.,  {Jackson} J.~M.,  {Kobulnicky} H.~A.,    {Mathis} J.~S.,
   2006, \apj, 649, 759

\bibitem[\protect\citeauthoryear{{Churchwell}, {Watson}, {Povich}, {Taylor},
  {Babler}, {Meade}, {Benjamin}, {Indebetouw} \& {Whitney}}{{Churchwell}
  et~al.}{2007}]{2007ApJ...670..428C}
{Churchwell} E.,  {Watson} D.~F.,  {Povich} M.~S.,  {Taylor} M.~G.,  {Babler}
  B.~L.,  {Meade} M.~R.,  {Benjamin} R.~A.,  {Indebetouw} R.,    {Whitney}
  B.~A.,  2007, \apj, 670, 428

\bibitem[\protect\citeauthoryear{{Dale}, {W{\"u}nsch}, {Whitworth} \& {Palou{\v
  s}}}{{Dale} et~al.}{2009}]{2009MNRAS.398.1537D}
{Dale} J.~E.,  {W{\"u}nsch} R.,  {Whitworth} A.,    {Palou{\v s}} J.,  2009,
  \mnras, 398, 1537

\bibitem[\protect\citeauthoryear{{Dawson}, {Kawamura}, {Mizuno}, {Onishi} \&
  {Fukui}}{{Dawson} et~al.}{2008}]{2008PASJ...60.1297D}
{Dawson} J.~R.,  {Kawamura} A.,  {Mizuno} N.,  {Onishi} T.,    {Fukui} Y.,
  2008, \pasj, 60, 1297

\bibitem[\protect\citeauthoryear{{Deharveng}, {Lefloch}, {Kurtz}, {Nadeau},
  {Pomar{\`e}s}, {Caplan} \& {Zavagno}}{{Deharveng}
  et~al.}{2008}]{2008A&A...482..585D}
{Deharveng} L.,  {Lefloch} B.,  {Kurtz} S.,  {Nadeau} D.,  {Pomar{\`e}s} M.,
  {Caplan} J.,    {Zavagno} A.,  2008, \aap, 482, 585

\bibitem[\protect\citeauthoryear{{Deharveng}, {Lefloch}, {Massi}, {Brand},
  {Kurtz}, {Zavagno} \& {Caplan}}{{Deharveng}
  et~al.}{2006}]{2006A&A...458..191D}
{Deharveng} L.,  {Lefloch} B.,  {Massi} F.,  {Brand} J.,  {Kurtz} S.,
  {Zavagno} A.,    {Caplan} J.,  2006, \aap, 458, 191

\bibitem[\protect\citeauthoryear{{Deharveng}, {Lefloch}, {Zavagno}, {Caplan},
  {Whitworth}, {Nadeau} \& {Mart{\'{\i}}n}}{{Deharveng}
  et~al.}{2003}]{2003A&A...408L..25D}
{Deharveng} L.,  {Lefloch} B.,  {Zavagno} A.,  {Caplan} J.,  {Whitworth} A.~P.,
   {Nadeau} D.,    {Mart{\'{\i}}n} S.,  2003, \aap, 408, L25

\bibitem[\protect\citeauthoryear{{Deharveng}, {Zavagno} \&
  {Caplan}}{{Deharveng} et~al.}{2005}]{2005A&A...433..565D}
{Deharveng} L.,  {Zavagno} A.,    {Caplan} J.,  2005, \aap, 433, 565

\bibitem[\protect\citeauthoryear{{Deharveng}, {Zavagno}, {Schuller}, {Caplan},
  {Pomar{\`e}s} \& {De Breuck}}{{Deharveng} et~al.}{2009}]{2009A&A...496..177D}
{Deharveng} L.,  {Zavagno} A.,  {Schuller} F.,  {Caplan} J.,  {Pomar{\`e}s} M.,
     {De Breuck} C.,  2009, \aap, 496, 177

\bibitem[\protect\citeauthoryear{{Ehlerov{\'a}} \& {Palou{\v
  s}}}{{Ehlerov{\'a}} \& {Palou{\v s}}}{2005}]{2005A&A...437..101E}
{Ehlerov{\'a}} S.,  {Palou{\v s}} J.,  2005, \aap, 437, 101

\bibitem[\protect\citeauthoryear{{Elmegreen}}{{Elmegreen}}{1994}]{1994ApJ...42%
7..384E}
{Elmegreen} B.~G.,  1994, \apj, 427, 384

\bibitem[\protect\citeauthoryear{{Elmegreen}}{{Elmegreen}}{1998}]{1998ASPC..14%
8..150E}
{Elmegreen} B.~G.,  1998, in {C.~E.~Woodward, J.~M.~Shull, \& H.~A.~Thronson
  Jr.} ed., Origins Vol.~148 of Astronomical Society of the Pacific Conference
  Series, {Observations and Theory of Dynamical Triggers for Star Formation}.
pp 150--+

\bibitem[\protect\citeauthoryear{{Elmegreen} \& {Lada}}{{Elmegreen} \&
  {Lada}}{1977}]{1977ApJ...214..725E}
{Elmegreen} B.~G.,  {Lada} C.~J.,  1977, \apj, 214, 725

\bibitem[\protect\citeauthoryear{{Fukui}, {Onishi}, {Abe}, {Kawamura},
  {Tachihara}, {Yamaguchi}, {Mizuno} \& {Ogawa}}{{Fukui}
  et~al.}{1999}]{1999PASJ...51..751F}
{Fukui} Y.,  {Onishi} T.,  {Abe} R.,  {Kawamura} A.,  {Tachihara} K.,
  {Yamaguchi} R.,  {Mizuno} A.,    {Ogawa} H.,  1999, \pasj, 51, 751

\bibitem[\protect\citeauthoryear{{Haffner}, {Reynolds}, {Tufte}, {Madsen},
  {Jaehnig} \& {Percival}}{{Haffner} et~al.}{2003}]{2003ApJS..149..405H}
{Haffner} L.~M.,  {Reynolds} R.~J.,  {Tufte} S.~L.,  {Madsen} G.~J.,  {Jaehnig}
  K.~P.,    {Percival} J.~W.,  2003, \apjs, 149, 405

\bibitem[\protect\citeauthoryear{{Hatzidimitriou}, {Stanimirovic},
  {Maragoudaki}, {Staveley-Smith}, {Dapergolas} \&
  {Bratsolis}}{{Hatzidimitriou} et~al.}{2005}]{2005MNRAS.360.1171H}
{Hatzidimitriou} D.,  {Stanimirovic} S.,  {Maragoudaki} F.,  {Staveley-Smith}
  L.,  {Dapergolas} A.,    {Bratsolis} E.,  2005, \mnras, 360, 1171

\bibitem[\protect\citeauthoryear{{Kim}, {Staveley-Smith}, {Dopita}, {Freeman},
  {Sault}, {Kesteven} \& {McConnell}}{{Kim} et~al.}{1998}]{1998ApJ...503..674K}
{Kim} S.,  {Staveley-Smith} L.,  {Dopita} M.~A.,  {Freeman} K.~C.,  {Sault}
  R.~J.,  {Kesteven} M.~J.,    {McConnell} D.,  1998, \apj, 503, 674

\bibitem[\protect\citeauthoryear{{McClure-Griffiths}, {Dickey}, {Gaensler} \&
  {Green}}{{McClure-Griffiths} et~al.}{2002}]{2002ApJ...578..176M}
{McClure-Griffiths} N.~M.,  {Dickey} J.~M.,  {Gaensler} B.~M.,    {Green}
  A.~J.,  2002, \apj, 578, 176

\bibitem[\protect\citeauthoryear{{Vishniac}}{{Vishniac}}{1983}]{1983ApJ...274.%
.152V}
{Vishniac} E.~T.,  1983, \apj, 274, 152

\bibitem[\protect\citeauthoryear{{Watson}, {Corn}, {Churchwell}, {Babler},
  {Povich}, {Meade} \& {Whitney}}{{Watson} et~al.}{2009}]{2009ApJ...694..546W}
{Watson} C.,  {Corn} T.,  {Churchwell} E.~B.,  {Babler} B.~L.,  {Povich} M.~S.,
   {Meade} M.~R.,    {Whitney} B.~A.,  2009, \apj, 694, 546

\bibitem[\protect\citeauthoryear{{Watson}, {Povich}, {Churchwell}, {Babler},
  {Chunev}, {Hoare}, {Indebetouw}, {Meade}, {Robitaille} \& {Whitney}}{{Watson}
  et~al.}{2008}]{2008ApJ...681.1341W}
{Watson} C.,  {Povich} M.~S.,  {Churchwell} E.~B.,  {Babler} B.~L.,  {Chunev}
  G.,  {Hoare} M.,  {Indebetouw} R.,  {Meade} M.~R.,  {Robitaille} T.~P.,
  {Whitney} B.~A.,  2008, \apj, 681, 1341

\bibitem[\protect\citeauthoryear{{Whitworth}, {Bhattal}, {Chapman}, {Disney} \&
  {Turner}}{{Whitworth} et~al.}{1994}]{1994MNRAS.268..291W}
{Whitworth} A.~P.,  {Bhattal} A.~S.,  {Chapman} S.~J.,  {Disney} M.~J.,
  {Turner} J.~A.,  1994, \mnras, 268, 291

\bibitem[\protect\citeauthoryear{{Whitworth} \& {Francis}}{{Whitworth} \&
  {Francis}}{2002}]{2002MNRAS.329..641W}
{Whitworth} A.~P.,  {Francis} N.,  2002, \mnras, 329, 641

\bibitem[\protect\citeauthoryear{{W{\"u}nsch} \& {Palou{\v s}}}{{W{\"u}nsch} \&
  {Palou{\v s}}}{2001}]{2001A&A...374..746W}
{W{\"u}nsch} R.,  {Palou{\v s}} J.,  2001, \aap, 374, 746

\bibitem[\protect\citeauthoryear{{Zavagno}, {Deharveng}, {Comer{\'o}n},
  {Brand}, {Massi}, {Caplan} \& {Russeil}}{{Zavagno}
  et~al.}{2006}]{2006A&A...446..171Z}
{Zavagno} A.,  {Deharveng} L.,  {Comer{\'o}n} F.,  {Brand} J.,  {Massi} F.,
  {Caplan} J.,    {Russeil} D.,  2006, \aap, 446, 171

\end{thebibliography}
\label{lastpage}
\end{document}